\documentclass[pra,twocolumn,showpacs,preprintnumbers,amsmath,amssymb,tightenlines,epsfig,superscriptaddress]{revtex4-1}
\usepackage{amsmath}
\usepackage{amssymb}
\usepackage{amsthm}
\usepackage{amsfonts}
\usepackage{graphicx}
\usepackage{bm}
\usepackage{xcolor}
\usepackage[percent]{overpic}
\usepackage{bbm}
\usepackage{lipsum}
 \usepackage[normalem]{ulem}

\newcommand{\bra}[1]{\langle\,{#1}\, |}
\newcommand{\ket}[1]{|\,{#1}\,\rangle}

\newcommand{\xvec}{\mathbf{x}}
\newcommand{\qvec}{\mathbf{q}}

\newcommand{\rvec}[1]{{\mathbf{r}}}

\newcommand{\comm}[2]{\big[#1,#2\big]}
\newcommand{\expec}[1]{\langle #1 \rangle}

\newcommand{\sub}[2]{{#1}_{\mbox{\!\! \scriptsize #2}}}

\newcommand{\bv}[1]{\mathbf{ #1 }}

\newcommand{\xv}{\mathbf{x}}
\newcommand{\qv}{\mathbf{q}}

\newcommand{\dv}{\mathbf{d}}

\def\beq{\begin{equation}}
\def\eeq{\end{equation}}

\def\CR{\nonumber\\[0.15cm]}

\newcommand{\fref}[1]{Fig.~\ref{#1}}
\newcommand{\frefp}[2]{Fig.~\ref{#1}~(#2)}

\newcommand{\eref}[1]{Eq.~(\ref{#1})}

\newcommand{\sref}[1]{section~\ref{#1}}

\newcommand{\cref}[1]{chapter~\ref{#1}}

\newcommand{\Cref}[1]{Chapter~\ref{#1}}
\newcommand{\tref}[1]{table~\ref{#1}}

\newcommand{\aref}[1]{appendix~\ref{#1}}
\newcommand{\bref}[1]{(\ref{#1})}

\usepackage{ulem}  
\begin{document}

\title{Tayloring Bose-Einstein condensate environments for a Rydberg impurity}
\author{S.~Rammohan}
\affiliation{Department of Physics, Indian Institute of Science Education and Research, Bhopal, Madhya Pradesh 462 023, India}
\email{sidharth16@iiserb.ac.in}
\author{A.~K.~Chauhan}
\affiliation{Department of Physics, Indian Institute of Science Education and Research, Bhopal, Madhya Pradesh 462 023, India}
\affiliation {Department of Optics, Faculty of Science, Palack\'y University, 17.~listopadu 1192/12, 77146 Olomouc, Czechia}
\author{R.~Nath}
\affiliation{Department of Physics, Indian Institute of Science Education and Research, Pune 411 008, India}
\author{A.~Eisfeld}
\affiliation{Max Planck Institute for the Physics of Complex Systems, N\"othnitzer Strasse 38, 01187 Dresden, Germany}
\author{S.~W\"uster}
\affiliation{Department of Physics, Indian Institute of Science Education and Research, Bhopal, Madhya Pradesh 462 023, India}
\email{sebastian@iiserb.ac.in}
\begin{abstract}
Experiments have demonstrated that the excitation of atoms embedded in a Bose-Einstein condensate to Rydberg states is accompanied by phonon creation. Here we provide the theoretical basis for the description of phonon-induced decoherence of the superposition of two different Rydberg states.
 To this end, we determine Rydberg-phonon coupling coefficients using a combination of analytical and numerical techniques. From these coefficients, we calculate bath correlation functions, spectral densities and re-organisation energies. These quantities characterize the influence of the environment and form essential inputs for follow-up open quantum system approaches. We find that the amplitude of bath correlations
scales like the power law $\nu^{-6}$ with the principal quantum number $\nu$, while re-organisation energies scale exponentially, reflecting the extreme tunability of Rydberg atomic properties.  
\end{abstract}

\maketitle

\section{Introduction} 
A growing arena in ultra-cold atomic physics is the study of impurities in quantum many body systems. Here a minority species composed of ions \cite{zipkes2010trapped,schmid2010dynamics,Kleinbach_Rydberg_ionimp,Sourav_ioncooling_PhysRevA,Dieterle_inelasticions_BEC_PRA}, different elements \cite{spethmann2012dynamics} or molecules \cite{wynar2000molecules} is embedded in a majority species that may form a Bose-Einstein condensate (BEC) \cite{schmid2010dynamics,Kleinbach_Rydberg_ionimp,klein2007dynamics}, thermal gas \cite{johnson2011impurity} or degenerate Fermi gas \cite{cetina2016ultrafast,schirotzek2009observation}. Besides the fundamental atomic physics interest, such experiments allow controlled tests of condensed matter impurity phenomena, ranging from the Kondo effect \cite{cyril1997kondo,Nakagawa_kondo_fermionlattice_PhysRevLett} over Polaron formation \cite{grusdt2017strong,camargo2018creation,schmidt2018theory,bruderer2007polaron,bruderer2008transport} to the Anderson orthogonality catastrophe \cite{Knap_imnmpurity_fermions_PhysRevX}.

While in the above examples the impurity typically is a particle in its electronic ground state, the impurity can also be a Rydberg excited atom of the same or another atomic species  as the cold gas \cite{heidemann2008rydberg,balewski:elecBEC,gaj:molspecdensshift,schlagmuller2016probing}. Rydberg atoms complement the above list of impurities in that they can interact equally strongly with a large but finite volume of the host medium, rather than dominantly with nearest neighbor atoms. The range and strength of interactions is further
highly controllable through the choice of the Rydberg quantum state. Being neutral, Rydberg atoms are not too sensitive to stray external fields, while they still can be guided by controlled external fields. However, being electronically excited, they suffer spontaneous decay. The resulting  lifetimes of tens or hundreds of microseconds \cite{beterov1989spontaneous,low2012experimental} are much shorter than the millisecond time-scales characteristic of BEC.
 We show in this article that 
interesting joint dynamics may arise between the two, despite this apparent time-scale mismatch.

\begin{figure}[htb]
\includegraphics[width=1\columnwidth]{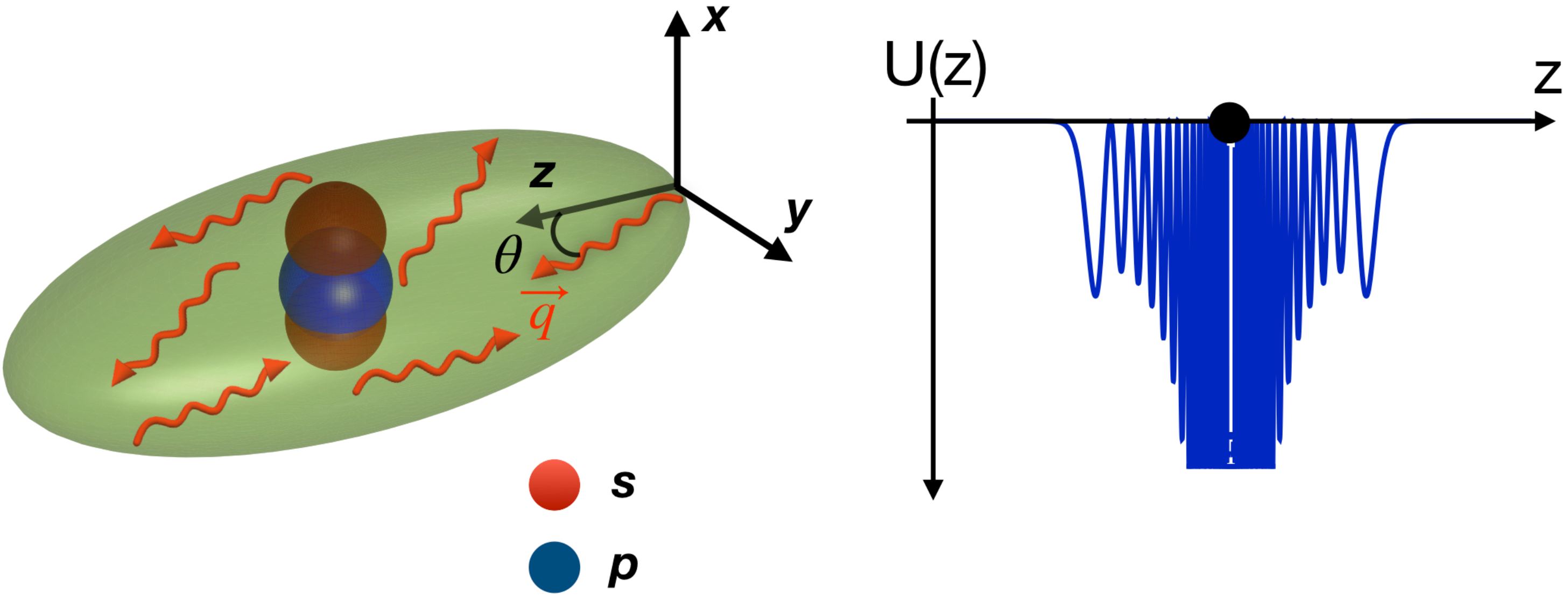}
\caption{\label{sketch} Sketch of a Rydberg impurity in a BEC environment. (left) A single Rydberg impurity in either of two selected internal states (s,p) couples to a large volume (blue or orange) of an embedding BEC (green), exciting phonons with wave-vector $\mathbf{q}$ (red arrows). The shape of the coupling volume strongly differs in the two Rydberg electronic states. (right) The system-environment coupling potential $U$ has a peculiar oscillatory long-range character.
 For illustration we sketch $U$ for an impurity in state $\ket{s}$, cut-off at large $|U|$ for better visibility, along an arbitrary axis, here $z$.
}
\end{figure}
We focus on a single Rydberg excited impurity atom embedded in a Bose-Einstein condensate. After re-writing the Hamiltonian in the form of a Spin-Boson-Model (SBM), we proceed to explicitly calculate the relevant Rydberg-phonon coupling constants. Focussing further on the case of two low lying angular momentum states ($l=0,1$), we infer bath correlation functions, spectral densities and re-organisation energies to characterise the phonon environment. It turns out the latter is highly tunable through Rydberg state quantum numbers and BEC phonon-mode structure. 

In our setup, the environment is naturally initialised in a coherent initial state, instead of the usual vacuum or thermal state. This happens due to the 
 sudden Rydberg excitation within a ground-state, zero-temperature BEC, resulting in a quench of the system.
A similar scenario is encountered for vibrational dynamics of molecules following photo-excitation \cite{Saikin_2013photonics}. Many open-quantum system techniques are formulated for environments in an initial vacuum or thermal state. These can still be used here since the dynamics for an environment in a coherent initial state can be mapped onto one for the environment in a vacuum initial state but adding auxiliary terms in the system Hamiltonian.  
 
Our results are validated by comparison with conceptually much simpler solutions of the Gross-Pitaveskii equation in a companion article \cite{rammohan:superpos}. We also show there, that the Rydberg in BEC system represents a particularly accessible example of an open quantum system where both, the system \emph{and} the environment can be interrogated in detail. The results may have further applications for the design of hybrid quantum technologies based on Rydberg atoms and BEC and creating flexible quantum simulation platforms for energy transport \cite{schonleber2015quantum,schempp:spintransport}.

BEC-phonon induced impurity decoherence has so far been mainly studied in the context of ground-state impurities of a minority species \cite{schmidt2018quantum,schmidt2019motional,song2019controlling,ostmann2017cooling,mcendoo2013entanglement,klein2007dynamics,cirone2009collective,bruderer2007polaron,bruderer2008transport,bruderer2006probing,lampo2018non,yuan2017quantum}, ions \cite{ratschbacher2013decoherence}, or polaron formation \cite{nielsen2019critical}.
For ground-state impurities, spectral densities were reported in \cite{haikka2013non}.

This article is organized as follows: Firstly we divide the many-body Hamiltonian in \sref{multi_species} into parts describing the system, the environment and the coupling between the two. Then in \sref{OQS}, within an open quantum system approach, we calculate environment correlation functions as well as spectral densities and then explore the scaling of the latter with principal quantum number. Finally we conclude along with an outlook in \sref{conclusion}. Details of calculations are provided in a set of appendices, with the incorporation of Bogoliubov excitations in the system environment coupling in \aref{app_ground_rydberg}, calculation of the ensuing coupling constants in \aref{app_kappas}, transformation of a coherent state environment into a vacuum one in \aref{app_coherentstate_treatment}, details on correlation functions in \aref{app_correlfct} and details on spectral densities in \aref{app_specdens}.

\section{Interacting multi-species system} 
\label{multi_species}

We begin with the many-body Hamiltonian $\hat{H}$ for a collection of Bosonic atoms of mass $m$. 
The internal states of the atoms are denoted by $k$. This label $k$ can for example correspond to the electronic ground-state $\ket{g}$ or to a collection of Rydberg states $\ket{\alpha}=\ket{\nu,l,m}$, with principal quantum number $\nu$, angular momentum $l$ and azimuthal quantum number $m$.

Using the field operator in the Heisenberg picture $\hat{\Psi}_k(\xv)$, which destroys an atom at location $\xv$ in internal state $k$, we have
\begin{align}
\hat{H}& = \sum_{k} \int d^3 \xvec\bigg[\hat{\Psi}^{\dagger}_k(\xvec)\Big(-\frac{\hbar^2}{2m} \boldsymbol{\nabla}^2+ E_k\Big)\hat{\Psi}_k(\xvec) 
\label{manybody_hamiltonian}
\\
& +\frac{1}{2}\sum_{i,j,s} \int d^3 \mathbf{y} \hat{\Psi}_k^{\dagger}(\xvec)\hat{\Psi}_i^{\dagger}(\mathbf{y})U_{kijs}(\xvec-\mathbf{y})\hat{\Psi}_j(\mathbf{y})\hat{\Psi}_s(\xvec)\bigg].
\nonumber
\end{align}
The first line of the Hamiltonian \bref{manybody_hamiltonian} are single particle energies: kinetic energy and internal electronic energies $E_k$. We do not consider any external potential.
The second term contains inter-atomic interactions, which may be long range due to the presence of Rydberg states and where we have allowed for interactions to change the internal state. 

We now focus on the scenario of a single Rydberg impurity that is allowed to occupy multiple electronic states, which is embedded in a majority BEC with atoms in the ground-state. Exploiting these constraints, we proceed in the following sub-sections to split the general Hamiltonian \bref{manybody_hamiltonian} into the three pieces that enter an open quantum system treatment \cite{Schlosshauer_decoherence_review,breuer_nonmarkovian_RevModPhys,book:maykuehn}, namely the sub-Hamiltonians for the quantum-system {(Rydberg atom states)}, the environment {(BEC)}, and the system-environment coupling, respectively:
\begin{align}
\hat{H}& =\sub{\hat{H}}{syst}+ \sub{\hat{H}}{env} + \sub{\hat{H}}{coup} .
\label{hamil_split}
\end{align}
Based on this segregation, we are able to evaluate the essential inputs for any open-quantum system approach, which are environment correlation functions or spectral densities.

\subsection{Rydberg quantum system}
\label{sec_quantsyst}

To make the above field operator notation compatible with the more usual formalism employed in Rydberg physics, we assume a highly localized Rydberg atom, restricting its position to a single, immobile spatial mode. We thus write
\begin{align}
\hat{\Psi}_\alpha(\mathbf{x}) \approx \varphi_0(\mathbf{x}) \hat{a}^{(\alpha)},
\label{rydberg_fieldop}
\end{align}
where $\hat{a}^{(\alpha)}$ creates a particle from the vacuum with internal state $\alpha$ and spatial mode $\varphi_0(\mathbf{x})$. For theRydberg atom multiple internal electronic states $\ket{\alpha}=\ket{\nu,l,m}$, defined above, are available. In the following we shall use the greek indices $\alpha$, $\beta$, with $\ket{\beta}=\ket{\nu ',l',m'}$ to refer to two such complete sets of quantum numbers.

For a single Rydberg impurity, with the identification $\hat{a}^{\dagger(\nu lm)}\hat{a}^{(\nu' l'm')}\leftrightarrow  \ket{\nu lm} \bra{\nu' l'm'}$, we thus reach the simple system Hamiltonian
\begin{align}
\sub{\hat{H}}{syst}& = \sum_{\nu l m} E_{\nu l m} \ket{\nu lm} \bra{\nu lm},
\label{Hsyst}
\end{align}
where $ E_{\nu lm}$ are the single atom energies corresponding to the state $\ket{\nu l m}$, which can be found with standard methods \cite{book:gallagher}. Since we assume a localized Rydberg atom $\varphi_0(\mathbf{x})\approx \delta^{(3)}(\mathbf{x}- \mathbf{R})$ at rest, we will ignore its kinetic energy operator in \eref{manybody_hamiltonian}.

We shall see later, that as usual the coupling to the BEC environment introduces energy shifts that are formally best included in $\sub{\hat{H}}{syst}$ as well, 
so that in \aref{final_interaction_Hamiltonian} we define a modified system Hamiltonian $\sub{\hat{H}}{syst}'$, see \eref{final_Hsyst}.

\subsection{Condensate environment}
\label{sec_environment}

For the ground-state atoms ($k=g$) that form the BEC, the Hamiltonian \bref{manybody_hamiltonian} becomes 
\begin{align}
\hat{H}& =  \int d^3\mathbf{x} \left[\hat{\Psi}^{\dagger}_g(\xvec)\Big(-\frac{\hbar^2}{2m} \boldsymbol{\nabla}^2+ E_g\Big)\hat{\Psi}_g(\xvec) \right.\nonumber\\
&  +\left. \frac{U_0}{2}\hat{\Psi}_g^{\dagger}(\xvec)\hat{\Psi}_g^{\dagger}(\xvec) \hat{\Psi}_g(\xvec)\hat{\Psi}_g(\xvec)\right],
\label{manybody_hamiltonian_gatoms}
\end{align}
assuming the usual s-wave contact interactions \cite{book:pethik}
\begin{align}
U_{gggg}(\xv - \mathbf{y})&=U_0 \delta^{(3)} (\xv - \mathbf{y}),
\label{Uint_groundstate}
\end{align}
with $\delta^{(3)}$ the three-dimensional delta-function and $U_0= 4\pi\hbar^2 a_s/m$, where $a_s$ is the s-wave atom-atom scattering length.

We split the ground-state field operator as usual~\cite{book:pethik}
\begin{equation}\label{gsfieldop}
\hat{\Psi}_g(\xv)=\phi_0(\xv)+\hat{\chi}(\xv),
\end{equation} 
where $\phi_0(\xv)\in{\mathbb C}$ is the mean-field condensate wave function and $\hat{\chi}(\xv)$ is the fluctuation operator which we expand as 
\begin{equation} \label{fluctop}
\hat{\chi}(\xv)=\sum_{\qv}\left(u_\qv(\xv)\hat{b}_\qv-v_\qv^*(\xv)\hat{b}^{\dagger}_\qv\right)
\end{equation}
in terms of Bogoliubov de-Gennes (BdG) excitations. In a homogenous BEC with number density $\rho=|\phi_0|^2$, these have mode functions $u_\qv(\xv)=\bar{u}_q\exp{[i \qv\cdot\xv]}/\sqrt{\cal V }$ and $v_\qv(\xv)=\bar{v}_q\exp{[i \qv\cdot\xv]}/\sqrt{\cal V }$ with bosonic creation and destruction operators $\hat{b}^{\dagger}_\qvec$ and $\hat{b}_\qvec$, assuming a box quantisation volume ${\cal V}$.
Here and in the following we use subscripts $q$ for quantities that only depend on the modulus of the quasi-particle wave number. In \bref{fluctop}, the BdG mode amplitudes are $\bar{u}_q=[(\zeta_q/\epsilon_q+1)/2]^{1/2}$ and $\bar{v}_q=[(\zeta_q/\epsilon_q-1)/2]^{1/2}$, with $\zeta_q=\epsilon_q+\rho U_0$, where $q=|\qvec|$ and
\begin{equation}  \label{BdGdisp}
\epsilon_q=\hbar\omega_q=\sqrt{\frac{\hbar^2q^2}{2m}\left(\frac{\hbar^2q^2}{2m}+2U_0\rho\right)}
\end{equation} 
are the BdG mode energies. The BdG amplitudes fulfil $\bar{u}_q^2-\bar{v}_q^2=1$ and $\lim_{q\rightarrow \infty }\bar{u}_q=1$, $\lim_{q\rightarrow \infty }\bar{v}_q=0$ with $\lim_{q\rightarrow 0} \bar{u}_q-\bar{v}_q=0$.

Inserting \bref{gsfieldop} and \bref{fluctop} into the Hamiltonian \bref{manybody_hamiltonian_gatoms} for the state $\ket{g}$ then as usual gives rise to the Hamiltonian rewritten in terms of quasiparticles
\begin{equation} 
\label{Hamil_g}
\sub{\hat{H}}{env}=E_{\rm GP}[\phi_0(\xvec)]+\sum_{\qv}\epsilon_q \hat{b}^{\dagger}_\qvec\hat{b}_\qvec,
\end{equation} 
where $E_{\rm GP}[\phi_0(\xvec)]$ is the Gross-Pitaevskii energy functional 
\begin{align} 
\!\!\!\!\!\!\!\!\!\!\!\! E_{GP}[\phi_0(\xvec)]&=\int d^3\xvec \bigg[-\frac{\hbar^2}{2m}|\boldsymbol{\nabla}\phi_0(\xvec)|^2\CR
&+E_g |\phi_0(\xvec)|^2+\frac{U_0}{2}|\phi_0(\xvec)|^4 \bigg].
 \label{EGP}
\end{align} 
%

\subsection{System-environment interactions}
\label{sec_interactions}

The main interest is in the system-environment coupling Hamiltonian $\sub{\hat{H}}{coup}$. 
Since we already dealt with interactions of ground-state atoms in \sref{sec_environment} and focus on at most one Rydberg excitation in the present article, the only remaining combinations of indices $kijs$ in the interaction part of the Hamiltonian \bref{manybody_hamiltonian} must involve one or two Rydberg indices only. We can exclude all terms that would involve transitions between ground and Rydberg states, due to the large energy difference and small wavefunction overlap of the impurity ground state wavefunction and the BEC. Considering these constraints, the only required index sets are $kijs$=$g\alpha\beta g$ and $kijs$=$\alpha g g\beta$, which describe the interaction of a ground-state with a Rydberg atom, possibly changing the internal state of the latter.

The dominant mechanism by which Rydberg atoms can interact with ground-state atoms, is through elastic scattering between the Rydberg electron and ground-state atoms, once the latter venture into the Rydberg orbit \cite{greene:ultralongrangemol}. This is described by the Fermi pseudopotential 
\begin{align}
\label{Fermi_potential}
V(\mathbf{y} + \mathbf{r},\xvec)  = g_0  \delta^{(3)}(\mathbf{y} + \mathbf{r} - \xvec),
\end{align}
where $g_0=2\pi\hbar^2 a_e/m_e$ \cite{omont1977theory}. Here, $a_e$ is the electron-atom scattering length with $a_e<0$ and $m_e$ the electron mass. We have split the absolute position of the Rydberg electron $\mathbf{y} + \mathbf{r}$ into the location of the ion core of the Rydberg atom $\mathbf{y}$, and the relative displacement of its electron $\mathbf{r}$. Interactions require the location of the electron to co-incide with that of a ground-state atom at $\mathbf{x}$. We neglect for simplicity the slight momentum dependence of the electron-atom scattering length $a_e$, see e.g.~\cite{eiles2019trilobites}, as well as the effect of the direct interaction with the ion core, which is relevant in a small BEC volume only \cite{schlagmuller2016single,flannery2005long}. The latter is also independent of electronic state, and hence not expected to contribute to decoherence.

In order to incorporate the potential \bref{Fermi_potential}, the Hamiltonian \bref{manybody_hamiltonian} could first be written down in terms of an explicit Rydberg electron position, as has been analyzed in \cite{middelkamp:rydinBEC}. Since energy differences between Rydberg states are much larger than the typical interaction energy scales with ground-state atoms, it is frequently useful to revert back to the atomic energy basis for the Rydberg electron, which we do here. Since we consider only a single impurity and $\comm{\hat{\Psi}_\alpha(\mathbf{x})}{\hat{\Psi}_\beta(\mathbf{y})}=0$, the system-environment interaction part of \bref{manybody_hamiltonian} finally boils down to
\begin{align}
\sub{\hat{H}}{int}& = \sum_{\alpha,\beta} \int d^3 \xvec \int d^3 \mathbf{y}\CR
 &\times \hat{\Psi}_g^{\dagger}(\xvec)\hat{\Psi}_\alpha^{\dagger}(\mathbf{y})U^{g\alpha\beta g}(\xvec-\mathbf{y})\hat{\Psi}_\beta(\mathbf{y})\hat{\Psi}_g(\xvec).
 \label{Hintg}
\end{align}
with long range ground-state Rydberg atom interaction 
\begin{align}
U^{g\alpha\beta g}(\textbf{x}-\textbf{y})=g_0\Big[\psi^{(\alpha)\:*}(\textbf{x}-\textbf{y})\psi^{(\beta)}(\textbf{x}-\textbf{y})\Big].
\label{Uint_rydrground}
\end{align}
Here $\psi^{(\alpha)}(\mathbf{r})$ denotes the electronic wave-function of the Rydberg electron in quantum state $\ket{\alpha}$ at a separation $\mathbf{r}$ from the core. The position $\mathbf{y}$ in \bref{Uint_rydrground} is that of the core of the Rydberg atom, and $\mathbf{x}$ that of a ground-state atom. For $\alpha=\beta$ the term \bref{Uint_rydrground} hence describes the energy shifts of Rydberg and ground-state atoms due to their proximity, while for $\alpha \neq \beta$ it allows the possibility that scattering from ground-state atoms causes a Rydberg state transition. 
The range of the interaction \bref{Uint_rydrground} is the extent of the Rydberg wavefunction $\psi$, which is slightly larger than the mean orbital radius $\sub{r}{orb}\approx 3a_0\nu^2/2$.

We now insert \eref{gsfieldop} and \bref{fluctop} into the ground state-Rydberg state interaction Hamiltonian \bref{Hintg} containing \bref{Uint_rydrground} and assume the core of the Rydberg atom to be very tightly localized in a single mode at the origin, using \eref{rydberg_fieldop}. We give some more details on intermediate steps, as well as some intial steps for an in-homogenous condensate in \aref{app_ground_rydberg}. 

After defining a splitting of the interaction Hamiltonian according to 
\begin{align}
\sub{\hat{H}}{int}&=\sum_{\alpha\beta} \hat{S}^{(\alpha\beta)} \otimes \hat{B}^{(\alpha\beta)} 
\label{Hamil_int}
\end{align} 
into system parts $\hat{S}^{(\alpha\beta)}=\hat{a}^{(\alpha)\dagger}\hat{a}^{(\beta)}$ and environment parts $\hat{B}^{(\alpha\beta)}$ and focussing on the case of a spatially homogeneous condensate with real mean field $\phi_0\approx\sqrt{\rho}$, where $\rho$ is the number density, we find
\begin{align}
 \hat{B}^{(\alpha\beta)} &=\Big[\bar{E}^{(\alpha\beta)}  +  \sum_\qv ( \kappa^{*(\alpha\beta)}_{\qv}\hat{b}^{\dagger}_{\qv} + \kappa^{(\alpha\beta)}_{\qv} \hat{b}_{\qv})\Big],  
\label{Hamil_int_E}
\end{align} 
with mean-field shift
\begin{align} 
\label{MFshift}
\bar{E}^{(\alpha\beta)}   &= g_0 \int d^3\xv  \:\: |\phi_0(\xv)|^2 \psi^{*(\alpha)}(\xv) \psi^{(\beta)}(\xv),
\end{align} 
and system-phonon coupling
\begin{align} 
\label{kappa_r_rprime}
\kappa^{(\alpha\beta)}_{\qv} &=g_0 \sqrt{\rho}  \int d^3\xv  \:\psi^{*(\alpha)}(\xv)\psi^{\beta}(\xv)[u_{\qv}(\xv)-v^*_{\qv}(\xv)].
\end{align} 
After incorporating \bref{MFshift} into $E_{\rm GP}[\phi_0(\xvec)]$, one can already use the Gross-Pitaevskii-equation (GPE) to study the mean-field dynamics of the condensate in the presence of a Rydberg impurity in a single state $\alpha$. One important effect in that case is imprinting of a phase onto the condensate wavefunction in any region where condensate atoms feel $U^{g\alpha\alpha g}$ \cite{mukherjee:phaseimp,Karpiuk_imaging_NJP}. This allows for example tracking of a mobile Rydberg impurity \cite{Tiwari_tracking} or distinguishing different electronic states \cite{Karpiuk_imaging_NJP}. In a homogenous system $\bar{E}^{(\alpha\beta)}=g_0\rho\,\delta_{\alpha,\beta}$, hence for the present purposes the term \bref{MFshift} only causes an inconsequential energy shift.

\begin{figure}[htb]
\includegraphics[width=\columnwidth]{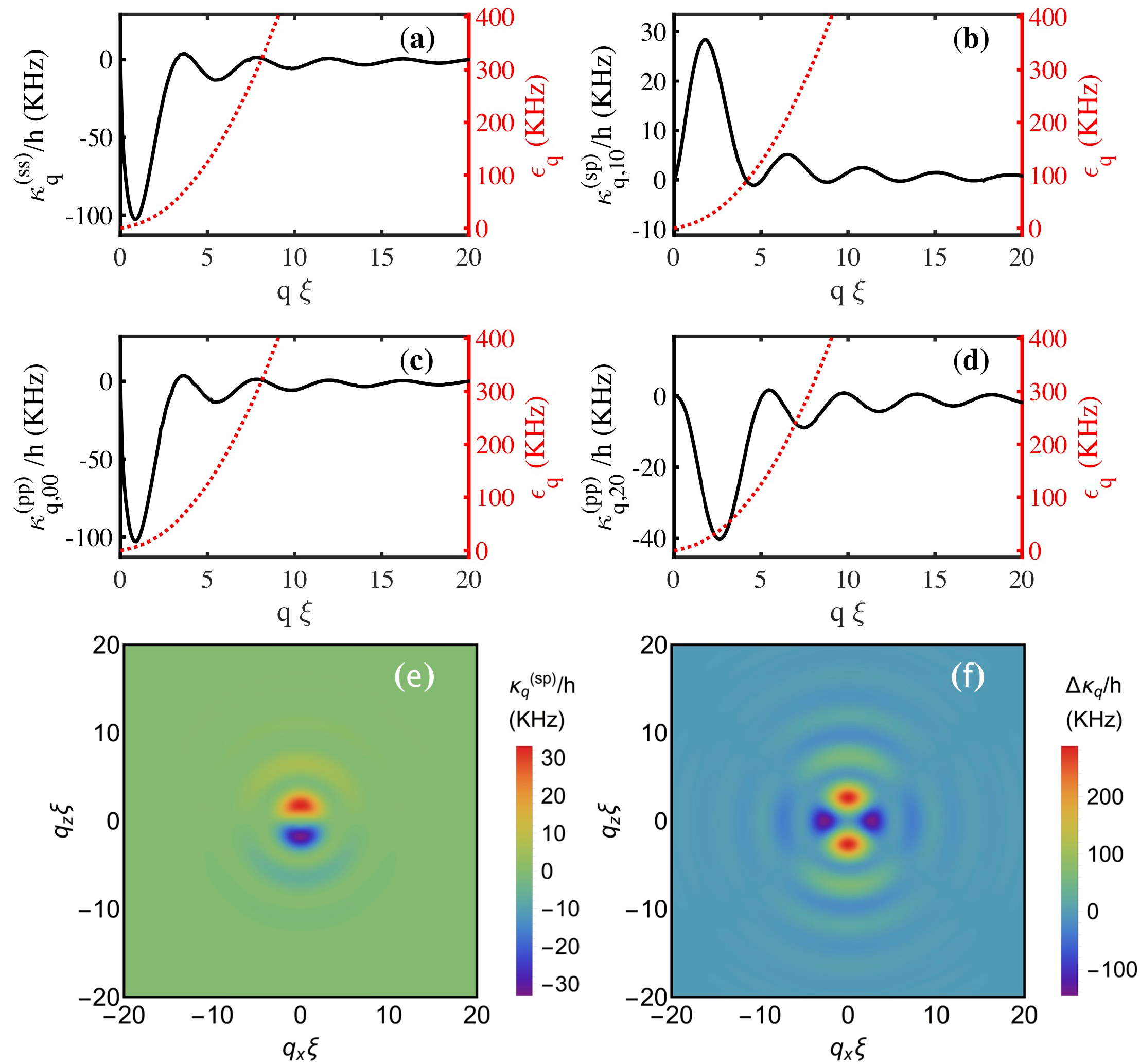}
\caption{\label{kappas} 
Overview of Rydberg-phonon coupling constants for a single impurity with $\nu=40$. (a) The coupling $\kappa^{(ss)}_q$ of state $\ket{s}$ to a phonon with wave-number $q$ is isotropic due to the symmetry of the Rydberg wave function. The right axis shows the phonon dispersion relation \bref{BdGdisp} for orientation.  (b) In contrast the transition coupling $\kappa^{(sp)}_{q,10}$ between $\ket{s}$ and $\ket{p}$ is anisotropic, since it involves $\ket{p}$ states. (c) Isotropic component $\kappa^{(pp)}_{q,00}$ of coupling in the state $\ket{p}$.
(d) Anisotropic component $\kappa^{(pp)}_{q,20}$ of coupling in the state $\ket{p}$.
(e) Transition coupling $\kappa^{(sp)}_q$ in the $q_x$, $q_z$ plane.
(f) The difference of couplings $\Delta\kappa_\qvec=\kappa^{(pp)}_q-\kappa^{(ss)}_q$ will be most relevant for Rydberg decoherence. 
}
\end{figure}

For $\alpha$, $\beta$ with $l=0$ and $l'=0$, denoted $(\alpha\beta)=(ss)$, the expression \bref{kappa_r_rprime} can be calculated analytically and has been used in \cite{balewski:elecBEC,balewski_thesis} to understand atom loss through repeated excitation of Rydberg impurities in a BEC; we list the result in \aref{app_kappas}. 
We approximate Rydberg electron wave functions by those for Hydrogen in the following, which should be a good approximation at large $\nu$. One finds that $\kappa^{(ss)}_{\qv}$ depends on $q=|\qv|$ only, with functional form shown in \frefp{kappas}{a}. For demonstrations in this section we consider the Rydberg impurity at the origin, in a $^{84}$Sr condensate with density $\rho=4.9\times10^{20}$ $\mathrm{m}^{-3}$, hence the relevant atom electron scattering length is $a_e=-18a_0$ \cite{bartschat2002ultralow}, the atomic mass $m=1.393\times10^{-25}$ kg and atom-atom scattering length $a_s=122.7a_0$ \cite{de2008two}.  The coupling coefficients still depend on the mode quantisation volume (we used ${\cal V}=204$ $\mu$m$^3$), defined after \eref{EGP}. ${\cal V}$ will drop out in subsequent results. We show the dependence on wavenumber multiplied by healing length $\xi=1/(2\sqrt{2\pi a_s\rho})$. Then $q \xi \lesssim 1$ corresponds to the phonon part of the BEC excitation spectrum and the remainder to the free particle one. For the parameters above, $\xi=0.11$ $\mu$m.

As discussed in the supplement of \cite{balewski:elecBEC}, the dominant peak in $\kappa^{(ss)}_{\qv}$, representing the largest value of the system-environment coupling, occurs at wave numbers determined by the size of the Rydberg orbit $q\approx 2/\sub{r}{orb}$ and lies in the regime of phonon excitations, as seen by comparison with the BdG dispersion relation shown as red dashed line. Further equally spaced peaks with alternating sign follow at wave numbers $q\geq1/\xi$ for which BEC excitations have single particle character.

When $l=1$ or $l'=1$, the integrand in \bref{kappa_r_rprime} is no longer isotropic. To be specific we restrict the present work to azimuthal quantum numbers $m=0$ assuming the quantisation axis along the $z$-direction. Removing other $m$ states from the picture can typically be achieved by additional Zeeman shifts through an external bias magnetic field~\cite{Ravets:angdep:PhysRevA.92.020701,leonhardt:orthogonal}.
The resultant coupling will then depend on the angle $\theta$ between the propagation direction of the phonon, $\bv{q}$, and the quantisation axis, sketched in \fref{sketch}. 
It turns out the angular dependence is described using just two spherical harmonics as $\kappa^{(\alpha\beta)}_{\qv}=\kappa^{(\alpha\beta)}_{q,00}(q) Y_{00} +\kappa^{(\alpha\beta)}_{q,20}(q) Y_{20}(\theta)$.

As described in \aref{app_kappas}, we perform the angular integration contained in \bref{kappa_r_rprime} analytically, the subsequent one over the radial variable $r=|\mathbf{x}|$ numerically. The results of this procedure are shown in \fref{kappas}.
We see as in the case of $\kappa^{(ss)}_{\qv}$, that the coupling to condensate excitations when $p$-states are involved has an oscillatory dependence on the excitation wave-number and extends over both, the phonon and the single particle part of the spectrum. We will see shortly, that the difference between couplings in the $s$ and the $p$ state, shown in panel (f), will be most relevant for studies of decoherence, since it encapsulates the ability of the BEC environment to ``measure'' the electronic state of the Rydberg system.

\section{Open quantum system approach}
\label{OQS}

If one is not interested in all degrees of freedom of a complex quantum system, it is convenient to split it into a system ${\cal S}$ and an environment ${\cal E}$, and then investigate the quantum dynamics of the system only. Formally the latter is given by
\begin{align}
\label{formal_reddm_evolution}
\hat{\rho}_{\cal S}(t) &=\mbox{Tr}_{\cal E} \left(\hat{U}(t)   \left[\hat{\rho}_{\cal S}(0)\otimes \hat{\rho}_{\cal E}(0)\right] \hat{U}^\dagger(t)\right),
\end{align}
where $\hat{\rho}_{\cal S}(t)$ is the reduced density matrix of the system, $\hat{U}(t)$ is the time evolution operator of the complete system and $\hat{\rho}_{{\cal S},{\cal E}}(0)$ are the initial reduced density matrices of system and environment, respectively. Since \bref{formal_reddm_evolution} is still based on the time-evolution operator of the complete system, it still contains the full complexity of the problem.

Open quantum system techniques aim to remove part of that complexity by finding an evolution equation that does not require to solve explicitly the bath degrees of freedom.
With the identification $\ket{\uparrow}=\ket{p}$ and  $\ket{\downarrow}=\ket{s}$ and constraining the Rydberg system to these two electronic states, we show in \aref{final_interaction_Hamiltonian} how the total Hamiltonian for our system can be rewritten as
\begin{align}
\label{total_system_Hamiltonian}
\hat{H}_{\text{tot}}=&\sub{\hat{H}}{syst}' + \sum_{\qv}\hbar\omega_q\:\tilde{b}^{\dagger}_{\qv}\tilde{b}_{\textbf{q}}+\sum_{\textbf{q}}\frac{\Delta\kappa_{\textbf{q}}}{2}\Big(\tilde{b}_{\textbf{q}}+\tilde{b}_{\textbf{q}}^{\dagger}\Big)\hat{\sigma}_z \nonumber \\
&+i\sum_\qvec \kappa_{\textbf{q}}^{(sp)}\Big(\tilde{b}_{\textbf{q}}-\tilde{b}_{\textbf{q}}^{\dagger}\Big)\hat{\sigma}_y+const.
\end{align}
In \bref{total_system_Hamiltonian} the environmental oscillator frequencies $\omega_q$ are set by the BdG mode energies in \bref{BdGdisp}. The coefficients $\kappa$ are defined in \eref{kappa_r_rprime} and the caption of \fref{kappas}, and $\hat{\sigma}_{y,z}$ are the usual Pauli spin operators. We recognize \bref{total_system_Hamiltonian} as a variant of the well-known Spin-Boson model (SBM) \cite{breuer2002theory,Leggett_dissipative_twolevel}.

The first term in \eref{total_system_Hamiltonian} is essentially \eref{Hsyst}, with minor energy shifts due to system-environment coupling discussed in \aref{final_interaction_Hamiltonian}. The BdG excitations in the second term created by $\tilde{b}^{\dagger}$ correspond to shifted harmonic oscillator modes, 
\begin{align}
\label{Bogoliubov_operators_coherent_bath}
\tilde{b}_{\textbf{q}}=\hat{b}_{\textbf{q}}+\frac{\bar{\kappa}_{\textbf{q}}}{2\hbar\omega_q}, \\
\tilde{b}_{\textbf{q}}^{\dagger}=\hat{b}_{\textbf{q}}^{\dagger}+\frac{\bar{\kappa}_{\textbf{q}}}{2\hbar\omega_q},
\end{align} 
with $\bar{\kappa}_{\textbf{q}}=\kappa_{\textbf{q}}^{(pp)}+\kappa_{\textbf{q}}^{(ss)}$, as discussed in detail in \aref{final_interaction_Hamiltonian}. We assume a Bose-Einstein condensate at temperature $T=0$ as initial state prior to Rydberg excitation, hence $\hat{\rho}_{\cal E}(0)=\ket{0}\bra{0}$, where $\ket{0}$ is the Bogoliubov vacuum of the original unshifted quasi-particle operators: $\hat{b}_{\textbf{q}}\ket{0}=0$. For the operators $\tilde{b}_{\textbf{q}}$ the initial state $\hat{\rho}_{\cal E}(0)$ corresponds to a many-mode coherent state as shown in \aref{final_interaction_Hamiltonian}.
This seems to be an obstacle for the application of open quantum system methods which require the environment initially to be in the vacuum state.
 However, as we demonstrate in \aref{app_coherentstate_treatment} the evolution according to \bref{formal_reddm_evolution} from the coherent initial state is equivalent to evolution from a vacuum initial state with a modified time-dependent system Hamiltonian
\begin{align}
\label{Hsyst_tot}
\sub{\hat{H}}{syst}''&=\left[ \frac{\Delta E}{2} + \sum_{\qv}\frac{\Delta\kappa_{\qv}\:\bar{\kappa}_{\qv}}{2\hbar\omega_q}\big(\cos(\omega_qt)-1\big) \right]\hat{\sigma}_z,
\end{align} 
where $\Delta E=E_p-E_s$ is the energy splitting between the $\ket{p}$ and the $\ket{s}$ state. For this we have combined \eref{Hsyst}, \bref{final_Hsyst} and \bref{shift_piece_hamil}.
Due to the equivalence discussed above, we shall thus consider the environment in the vacuum state also for the shifted operators $\tilde{b}_{\textbf{q}}$, but using \bref{Hsyst_tot} for the system.

Finally, for later use, let us reformulate the interaction Hamiltonian within \bref{total_system_Hamiltonian} as 
\begin{align}
\label{Hint_revamp}
\hat{H}_{\text{int}}&=\hat{\sigma}^{(z)}\otimes\hat{E}^{(z)}+\hat{\sigma}^{(y)}\otimes\hat{E}^{(y)}, \\
\hat{E}^{(z)}&=\sum_{\textbf{q}}\frac{\Delta\kappa_{\textbf{q}}}{2}\Big(\tilde{b}_{\textbf{q}}+\tilde{b}_{\textbf{q}}^{\dagger}\Big)\label{Ez_operator},
\\
\hat{E}^{(y)}&=i \sum_\qvec  \kappa_{\textbf{q}}^{(sp)}\Big(\tilde{b}_{\textbf{q}}-\tilde{b}_{\textbf{q}}^{\dagger}\Big).
\label{Ey_operator}
\end{align}
We shall in the following work out environmental properties such as correlation functions in terms of these new operators $\hat{E}^{(z)}$ and $\hat{E}^{(y)}$. 
As usual, the system environment interaction Hamiltonian leads to entanglement between system and environment and thus ultimately to decoherence.

\subsection{Single impurity Bath correlation functions}
\label{correlfct}
%
In an approximate reduced description for the system, all effects of the environment can usually be taken into account through the environment correlation functions
 \begin{equation} 
C^{(kl)}(\tau)=\bra{0} \hat{E}^{(k)}(\tau) \hat{E}^{(l)}(0) \ket{0},
 \label{C_of_tau_vac}
 \end{equation} 
where $k,l\in\{y,z\}$, operators are understood in the interaction picture and $\tilde{b}_{\textbf{q}} \ket{0}=0$ as discussed above. 

\begin{figure}[htb]
\includegraphics[width=1\columnwidth]{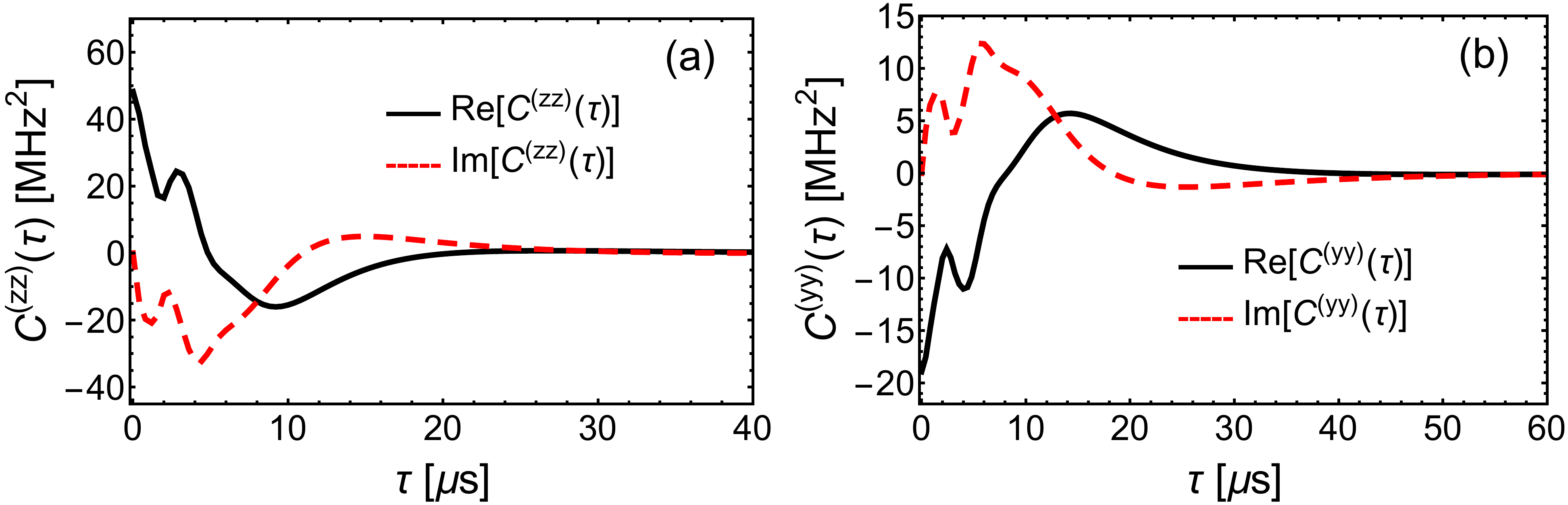}
\caption{\label{fig_correlation_functions} Phonon correlation functions, defined in \eref{C_of_tau_vac}, for coupling to a Rydberg impurity with principal quantum number $\nu=40$. (a) Re$[C^{(zz)}(\tau)]$ (black) and Im$[C^{(zz)}(\tau)]$  (red-dashed), (b) $C^{(yy)}(\tau)$ with the same line-styles. Other correlations vanish: $C^{(zy)}(\tau)=C^{(yz)}(\tau)=0$.
}
\end{figure}
We show in \fref{fig_correlation_functions} the relevant correlation functions for the two system-bath coupling operators $\hat{E}^{(z)}$, $\hat{E}^{(y)}$ in \bref{total_system_Hamiltonian}, for the same parameters as in \fref{kappas}. The third possible correlation function $C^{(zy)}(\tau)=\expec{ \hat{E}^{(z)}(\tau) \hat{E}^{(y)}(0)}$ vanishes. The calculations are described in more detail in \aref{app_correlfct}. We can already estimate typical decoherence time-scales $\sub{T}{dc}$ for Rydberg electronic state superpositions from these results, according to $\sub{T}{dc}\sim1/\sqrt{2C^{(zz)}(0)}$ \cite{book:schlosshauer}. For the case of \fref{fig_correlation_functions} we obtain $\sub{T}{dc}\approx 20$ ns. This is much shorter than the lifetime of $\nu=40$ Rydberg states, about $\tau\approx40$ $\mu$s \cite{beterov:BBR}, or decoherence times in vacuum, which can be of the order of milli-seconds even near surfaces \cite{Hermann_Avigliano_Rydbergcoherence_PRArapid}.

Another important aspect visible in \fref{fig_correlation_functions} are the phonon environment memory times $T_m$, over which correlation functions drop to zero. 
Let us loosely refer to the characteristic time-scale of the Rydberg impurity as $\sub{T}{sys}$. This can be either given by the energy splitting between our two states $\sub{T}{sys}\sim h/|E_\alpha - E_\beta|$, or if these are coupled by a micro-wave set by its Rabi-frequency: $\sub{T}{sys}\sim h/\sub{\Omega}{mw}$. Then, for characteristic Rydberg system time-scales $\sub{T}{sys}\gg T_m$ we would expect Markovian open quantum system dynamics, for $\sub{T}{sys} \leq T_m$ non-Markovian. We can read off from \fref{fig_correlation_functions} that $T_m\approx 50 \mu$s at $\nu=40$. Dynamics of either kind discussed above can typically be faster, hence we expect our system to be able to show non-Markovian features. It will in fact be difficulty to generate a non-trivially evolving quantum system that is Markovian, since the Rydberg evolution time would be limited by the radiative life-time, which even in vaccuum is of the same order as $T_m$.

By inspecting the Hamiltonian \bref{total_system_Hamiltonian}, we see that $C^{(yy)}$ is related to phonon induced transitions between Rydberg states. One would expect those to be strongly suppressed for phonon energies in the kHz range, and Rydberg energy splittings of GHz for energetic reasons. We confirm this expectation in \cite{rammohan:superpos}.

\subsection{Phonon spectral densities and environment tuning}
\label{spec_dens_tuning}

To isolate temperature effects that are encoded in the environment initial state $\hat{\rho}_{\cal E}(0)$, from the features of the system-environment coupling, one also frequently considers the environment  spectral density $J(\omega)$ that encapsulates the relevance of environmental degrees of freedom with frequency $\omega$. In our case, spectral densities are defined as
\begin{align} 
J^{(z)}(\omega) &= \sum_\qvec\frac{\Delta\kappa_{\qvec}^2}{4}\delta(\omega -\omega_\qvec), \CR
J^{(y)}(\omega) &= \sum_\qvec \kappa^{(sp)2}_{\qvec}\delta(\omega -\omega_\qvec).
 \label{specden}
\end{align} 
Since our environment is in the vacuum state, these can also be written as the Fourier transform of the bath correlation functions in \sref{correlfct}, via: 
\begin{align}
J^{(z)}(\omega)=\frac{1}{2\pi}\int_0^{\infty} d \tau\:C^{(zz)}(\tau)e^{i\omega\tau}, \CR
J^{(y)}(\omega)=\frac{1}{2\pi}\int_0^{\infty} d \tau\:C^{(yy)}(\tau)e^{i\omega\tau}.
\label{JfromC}
\end{align}
\begin{figure}[htb]
\includegraphics[width=1\columnwidth]{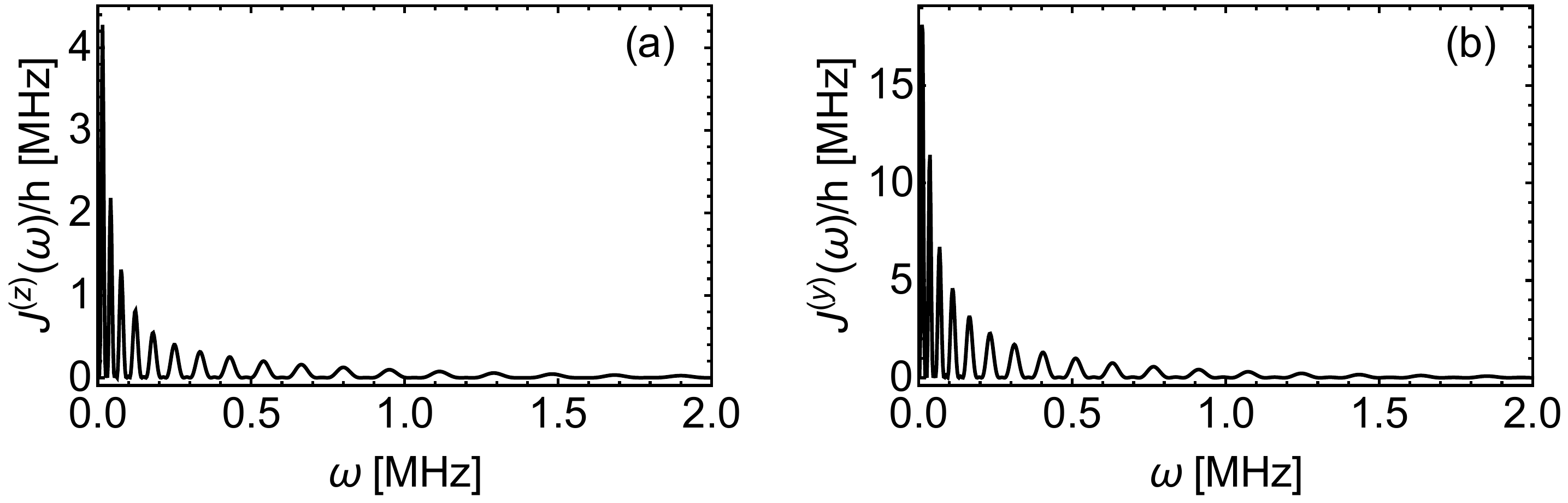}
\caption{\label{specdens} Spectral properties of BEC phonon environment for a Rydberg impurity in $\nu=120$. (a) Spectral density $J^{(z)}(\omega)$ and (b) $J^{(y)}(\omega)$. These are defined by \eref{specden} or equivalently \eref{JfromC}. 
}
\end{figure}
The results are shown in \fref{specdens}, with details of the calculation in \aref{app_specdens}. The spectral densities display a non-trival series of peaks and thus indicate the presence of a \emph{structured environment}. The structure originates from that of the coupling constants $\kappa_\qvec$ in \fref{kappas}, since the spectral densities are found as 
Fourier transform of the Bath correlation functions $C$. These in turn are a type of inverse Fourier transform of the $\kappa_\qvec^2$ according to \eref{Correlation_zz} and \eref{Correlation_yy}. However since one transform is in terms of the variable pair $(\omega,\tau)$ and the other in term of $(\qvec,\tau)$, the peaks are now no longer equidistant, but stretched in accordance with the dispersion relation $\omega(\qvec)$ \bref{BdGdisp}.

\begin{figure}[htb]
\includegraphics[width=1\columnwidth]{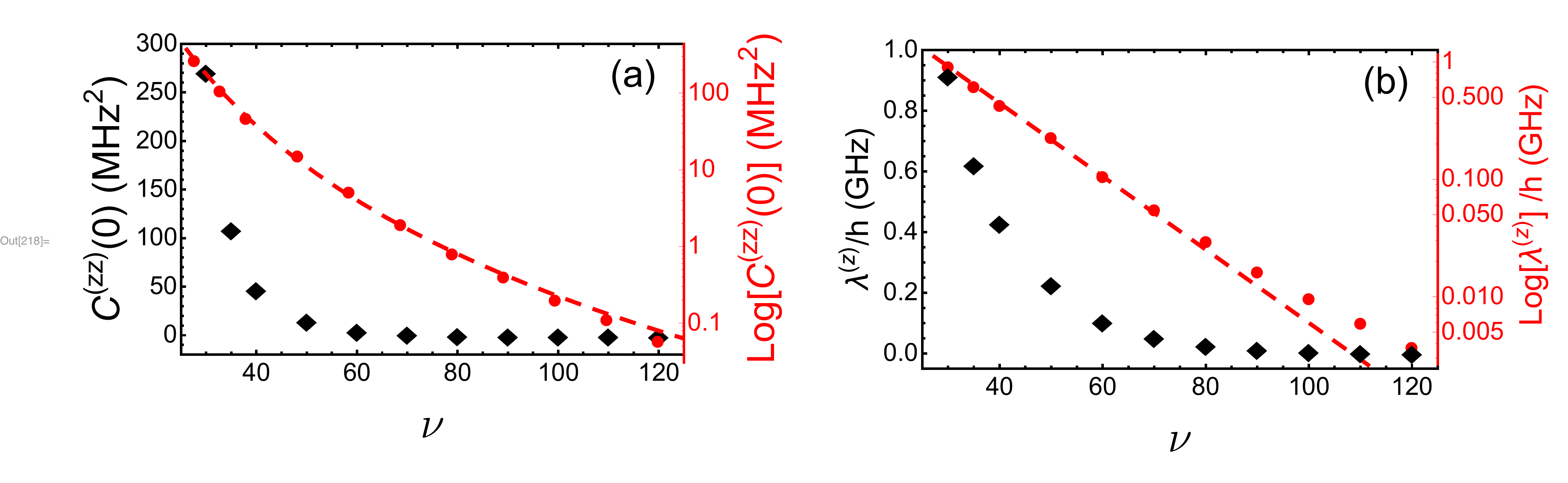}
\caption{\label{tuning} Tuning of the condensate environment. (a) (black diamonds) Variation of the bath auto-correlation function $C^{(zz)}(0)$ as a function of principal quantum number $\nu$ along with the power-law fit $C^{(zz)}(0)=1.25\times10^{11}\: \nu^{-6}$ MHz$^2$ as a red-dashed line. 
We show both, a linear (black) and a logarithmic axis (red). (b) The re-organisation energy $\lambda^{(z)}$, see \eref{reorgaz}, with the exponential fit $\lambda^{(z)}=5.76\times10^3\: e^{-0.06\times\nu}$ GHz.
}. 
\end{figure}

Either spectral densities such as in \fref{specdens} or bath correlation functions as in \fref{fig_correlation_functions} now fully capture the effect of the condensate environment on the Rydberg impurity.
Note that the bath correlation function at time zero, which entered the estimation of the decoherence time, is directly related to the integral over the corresponding spectral density.
 Given the extreme scaling of Rydberg electronic state properties with principal quantum number, we now expect a similar degree of tunability in the influence of the environment.
To demonstrate that this is indeed the case, \fref{tuning} shows how two measures for the impact of the environment on the system depend on the Rydberg principal quantum number $\nu$. The first measure is the initial value of the bath correlation function $C(0)$, shown in \frefp{tuning}{a}. Another frequently used measure is the re-organisation energy 
\begin{equation} 
\lambda^{(z)} =  \int_0^\infty \frac{J^{(z)}(\omega)}{\omega} d\omega,
 \label{reorgaz}
\end{equation} 
which is shown in \frefp{tuning}{b}. Large values for either quantity indicate a fast decohering effect of the environment.
We can see that system-environment coupling can be tuned over orders of magnitude through the principal quantum number.
Despite the wider excursions of the Rydberg electron into the ambient BEC medium for the higher principal quantum numbers, we find stronger system bath coupling at lower principal quantum numbers, because the stronger confinement of the Rydberg wavefunction at lower $\nu$ leads to a higher electron probability density, which in turn determines the coupling strength.

While we have focussed here on the tuning of system-environment coupling through choice of Rydberg properties, an alternate route is a modification of the ambient condensate. We can see from \eref{kappa_r_rprime} in conjunction with \bref{C_of_tau_vac} or \bref{specden}, that all quantities in \fref{tuning} are multiplied with the BEC density $\rho$. Additionally, the condensate mean-field interaction strength $U_0$ and density enter the expression through the phonon energies $\epsilon_\qvec$ in \bref{BdGdisp} and BdG mode amplitudes. Interactions $U_0$ can be tuned using Feshbach resonances \cite{book:pethik}, thus both quantities will provide additional control knobs. 
Finally, when going beyond a homogenous system, the non-trivial spatial BdG mode shapes $u_\mathbf{q}(\mathbf{x})$ and $v_\mathbf{q}(\mathbf{x})$ entering \eref{kappa_r_rprime} will depend on the system geometry, for example the trapping.  A final interesting aspect would be the scaling in angular momentum $l$, $l+1$ of the two states involved in  \fref{tuning}, where presently we fixed $l=0,1$. 

We defer explorations of the above options to the future.

\section{Conclusions and outlook} 
\label{conclusion}

We have studied a Rydberg excited atom within an atomic Bose-Einstein condensate, treating the latter as a controllable environment for the former. 
For this environment of phonon excitations in the BEC, we have introduced a framework for the calculation of bath-correlation functions, spectral densities and re-organisation energies.
These are important quantities for open quantum system approaches and contain all relevant information about the environmental influence on the systems dynamics. Knowledge of the re-organization energy or the bath correlation function at time zero already allows one to estimate the relevant time-scales for decoherence and phonon induced Rydberg state transitions. We found that they vary over orders of magnitude as a function of the principal quantum number $\nu$. 

For the example of $^{84}$Sr atoms, we find that estimated decoherence time-scales $\tau_\mathrm{decoh}=1/\sqrt{2C^{(zz)}(0)}$ between angular momentum states $\ket{\nu s}$ and $\ket{\nu p}$ range from $\sub{T}{dc}=5$ ns at a principal quantum number $\nu=40$ to $\tau_\mathrm{decoh}=0.9$ $\mu$s at $\nu=120$. We show in a companion article \cite{rammohan:superpos}, that phonon induced Rydberg state transitions between $\ket{\nu s}$ and $\ket{\nu p}$ are negligible on these time scales.  Bath memory times change from around $\sub{T}{m}=56$ $\mu$s at a principal quantum number $\nu=40$ to $\sub{T}{m}=660$ $\mu$s at  $\nu=120$. Since Rydberg dynamics is limited by lifetimes in a similar range, they would thus necessarily be placed in the non-Markovian regime where system dynamics happens on time-scales faster or comparable to the memory time.

While the resultant open-quantum system is already strongly tuneable in its system-environment coupling by variation of the principal quantum number of the Rydberg atom, additional tunability might arise, when extending the interaction model for example to long-range dressed interactions \cite{mukherjee:phaseimp}. By affecting also the condensate atoms \cite{nils:supersolids}, dressing additionally modifies the dispersion relation \bref{BdGdisp} which will in turn modify the spectral properties of the BEC environment.
Further control knobs would arise through the shape of the condensate wavefunction when moving from the homogenous condensate considered here, to tightly trapped clouds of atoms.

In the present work we have focused on the decoherence dynamics of the Rydberg atom. One might now also be interested in the coherence and correlation properties of the ambient BEC. So far schemes based on ground state impurities have been investigated and it has been shown theoretically that one can obtain information on coherence and correlation properties \cite{ng2008single,bruderer2006probing,streif2016measuring} and also on the temperature \cite{sabin2014impurities,mehboudi2019using} of the BEC. 
These schemes can now directly be adapted to the case of the Rydberg impurity studied in the present work, using the present results.
It is possible that Rydberg impurities offer advantages over ground-state atoms for some of these purposes, due to their stronger coupling to the condensate.

\acknowledgments
We gladly acknowledge fruitful discussions with Rick Mukherjee, and thank the Science and Engineering Research Board (SERB), Department of Science and Technology (DST), New Delhi, India, for financial support under research Project No.~EMR/2016/005462. We further are grateful for financial support from the Max-Planck society under the MPG-IISER partner group program. R.N.~acknowledges a UKIERI- UGC Thematic Partnership No. IND/CONT/G/16-17/73 UKIERI-UGC project. A.~E.~acknowledges support from the DFG via a Heisenberg fellowship (Grant No. EI 872/5-1).
\appendix
\section{Ground-Rydberg state interaction Hamiltonian}
\label{app_ground_rydberg} 
%
As discussed in \sref{sec_interactions}, the Hamiltonian for ground-state Rydberg-state interactions is
\begin{align}
\sub{\hat{H}}{int}& = \sum_{\alpha,\beta} \int d^3 \xvec \int d^3 \mathbf{y}\CR
 &\times \hat{\Psi}_g^{\dagger}(\xvec)\hat{\Psi}_\alpha^{\dagger}(\mathbf{y})U^{g\alpha\beta g}(\xvec-\mathbf{y})\hat{\Psi}_\beta(\mathbf{y})\hat{\Psi}_g(\xvec),
\label{interaction_hamiltonian_fieldop}
\end{align}
with $\hat{\Psi}_{g,\alpha,\beta}(\mathbf{x})$ field operators that destroy a ground state atom, or Rydberg atom in states $\ket{\alpha}$, $\ket{\beta}$ respectively, at position $\mathbf{x}$. 
Here $\alpha$ denotes a complete set of quantum numbers $\{\nu,l,m\}$, and similarly $\beta$ groups  $\{\nu',l',m'\}$.
We now insert the potential $U^{g\alpha\beta g}(\xvec-\mathbf{y})$ from \bref{Uint_rydrground} into \bref{interaction_hamiltonian_fieldop} and then assume the BdG expansion \bref{gsfieldop} for the ground-state field operator. Additionally using the Rydberg field operator \bref{rydberg_fieldop} restricted to a single mode $\hat{\Psi}_\alpha(\mathbf{x}) \approx \varphi_0(\mathbf{x}) \hat{a}^{(\alpha)}$, we obtain
\begin{align} 
\label{Rydberg_BEC_interaction_Hamiltonian_4}
\hat{H}_{\text{int}}=&g_0\sum_{\alpha,\beta}\int\int d^3\textbf{x}\hspace{1pt}d^3\textbf{y}\Big(\phi_0^*(\xv)+\hat{\chi}^{\dagger}(\xv)\Big)\varphi^*_0(\mathbf{y}) \hat{a}^\dagger_\alpha \CR
&\times \psi^{*(\alpha)}(\textbf{x}-\textbf{y}) \psi^{(\beta)}(\textbf{x}-\textbf{y})\CR
&\times\varphi_0(\mathbf{y})\hat{a}^\dagger_\beta \Big(\phi_0(\xv)+\hat{\chi}(\xv)\Big).
\end{align} 
So far, the single mode $\varphi_0(\mathbf{x})$ could still be de-localized, e.g.~a trap ground state.
 In the following we assume a fixed location of the Rydberg core at the origin, so that $|\varphi_0(\mathbf{y})|^2\approx \delta^{(3)}(\mathbf{y})$. Inserting the expansion \bref{fluctop} of the fluctuation operators into BdG modes, we then reach
\begin{align}
\label{Rydberg_BEC_interaction_Hamiltonian_5}
\hat{H}_{\text{int}}=& g_0\sum_{\alpha,\beta}\hat{a}^{\dagger(\alpha)} \hat{a}^{(\beta)} \int d^3\textbf{x} \bigg\{\:|\phi_0(\textbf{x})|^2 \CR
&+\bigg[\phi_0^*(\textbf{x})\sum_{\mathbf{q}}\left(u_{\mathbf{q}}(\textbf{x})\hat{b}_{\mathbf{q}}- v_{\mathbf{q}}^*(\textbf{x})\hat{b}^{\dagger}_{\mathbf{q}}\right)\CR
& + \phi_0(\textbf{x})\sum_{\mathbf{q}}\left(u_{\mathbf{q}}^*(\textbf{x})\hat{b}^{\dagger}_{\mathbf{q}}-v_{\mathbf{q}}(\textbf{x})\hat{b}_{\mathbf{q}}\right)\bigg] \bigg\} \CR
& \times \psi^{*(\alpha)}(\textbf{x})\psi^{(\beta)}(\textbf{x}).
\end{align}
As next simplification, we consider a real condensate mean field, $\phi_0(x)$=$\phi_0^*(x)$, which excludes for example condensates with non-trivial velocity profile.
It includes, however, the homogeneous static case treated here later, and typical simple trapped cases. Since we shall deal with a single impurity, we can finally identify $\hat{a}^{\dagger(\alpha)}\hat{a}^{(\beta)}$ with $\ket{\alpha}\bra{\beta}$, and cast \bref{Rydberg_BEC_interaction_Hamiltonian_5} into the form 
\begin{align}
\label{Hamiltonian_matrix_final}
\hat{H}_{\text{int}}=\sum_{\alpha,\beta}\ket{\alpha}\bra{\beta}\sum_{\mathbf{q}}\left[\kappa_{\mathbf{q}}^{(\alpha\beta)}\hat{b}_{\mathbf{q}}+\kappa_{\mathbf{q}}^{*(\alpha\beta)}\hat{b}_{\mathbf{q}}^{\dagger}+E^{(\alpha\beta)}\right],
\end{align}
with
\begin{align}
\label{coupling}
\kappa_{\mathbf{q}}^{(\alpha\beta)}&=g_0\int\:d^3\xv \: \psi^{*(\alpha)}(\textbf{x})\psi^{(\beta)}(\textbf{x})\phi_0(\textbf{x})\big( u_{\mathbf{q}}(\textbf{x})-\hspace{1pt}v_{\mathbf{q}}(\textbf{x})\big),\\
\label{E_alpha}
E^{(\alpha\beta)}&=g_0\int\:d^3\xv\: \psi^{*(\alpha)}(\textbf{x})\psi^{(\beta)}(\textbf{x})|\phi_0(\textbf{x})|^2,
\end{align}
which for a  homogenous condensate $\phi_0(\textbf{x})=\sqrt{\rho}$ with density $\rho$ reduce to the expressions given in \bref{MFshift} and \bref{kappa_r_rprime}.

\section{Calculation of coupling constants}
\label{app_kappas} 

The expression for coupling constants $\kappa^{(\alpha\beta)}_{\qv}$ in \eref{coupling} applies for a Rydberg atom in an arbitrary real condensate. We now consider the simpler homogeneous case, which should be a good approximation whenever the condensate density does not significantly vary on length scale of the Rydberg orbital radius $\sub{r}{orb}$. In that case the Bogoliubov modes take the simple plane wave form
\begin{align} 
\label{Bogoluibov_homog}
u_{\qv}(\xv)= \frac{\bar{u}_{q}}{\sqrt{\cal V}}e^{i{\qv}\cdot{\xv}}, \hspace{10pt}   v_{\qv}(\xv)=\frac{\bar{v}_{q}}{\sqrt{\cal V}}e^{i{\qv}\cdot{\xv}}, 
\end{align} 
using $q=|\qvec|$. Then \bref{coupling} becomes
\begin{align} 
\label{re-written_kappa}
\kappa^{(\alpha\beta)}_{\qv} &=\frac{g_0\sqrt{\rho}}{\sqrt{\cal V}}(\bar{u}_{q}-\bar{v}_{q}) \int d^3\xv  \:\:  \psi^{*\:(\alpha)}(\xv)\psi^{(\beta)}(\xv)e^{i\qv\cdot\xv}.
\end{align} 
We thus see that coupling constants are related to the Fourier transform of spatial Rydberg electron probability densities, for $\alpha=\beta$, or of products of two wave functions, for  $\alpha\neq\beta$. The prefactor $\bar{u}_{q}-\bar{v}_{q}\rightarrow 0 $ for $q\leq \xi$ and approaches one for $q> \xi$, where $\xi$ is the healing length.

At this stage we expand the plane waves in terms of spherical harmonics $Y_{lm}$, according to
\begin{align} 
\label{plane_wave_expansion}
e^{i {\qv}\cdot\mathbf{x}} &= 4 \pi \sum_{l_1=0}^{\infty}\:\sum_{m_1=-l_1}^{l_1} \:i^{l_1} j_{l_1}(qr) \: Y_{l_1m_1}(\hat{\mathbf{q}})\: Y_{l_1m_1}^*(\hat{\mathbf{x}}).
\end{align} 
Here $j_{l_1}(\text{qr})$ are spherical Bessel functions of the first kind, $\hat{\mathbf{q}}=\qvec/|\qvec|$ is a unit-vector along the wave-vector of the phonon, and 
$\hat{\mathbf{x}}=\mathbf{x}/|\mathbf{x}|$, while $r=|\mathbf{x}|$ and $q=|\mathbf{q}|$.

\subsection{Evaluation of angular integrals for general quantum states}
\label{angular_integral_general_states} 

As described before, the indices for electronic states $\alpha$ [$\beta$], are shorthand for quantum numbers $(n,l,m)$ [$(n',l',m')$].
The corresponding electronic wave-function in the Rydberg state of a Hydrogen or Alkali atom can be written as
\begin{align} 
\label{electronic_wavefunction}
\psi^{(\alpha)}\equiv\psi_{\nu lm}=\mathcal{N}_{\nu l} \:R_{\nu l}(r) Y_{lm}(\Theta,\Phi),
\end{align} 
with normalisation constant $\mathcal{N}_{\nu l}$ and radial wave function $R_{\nu l}(r)$, while the angular wave functions are spherical harmonics $Y_{lm}(\Theta,\Phi)$ in terms of angular coordinates of the electron $\Theta$ and $\Phi$. Note that we use capitalized angles for the coordinates of the electron, and lower case ones for the direction of the phonon wave vector $\qvec$.
For Hydrogen states that we use in the following, $\mathcal{N}_{\nu l}=\sqrt{\Big(\frac{2}{\nu a_0}\Big)^3\frac{(\nu-l-1)!}{2\nu[(\nu+l)!]}}$,
 with Bohr radius $a_0$. The radial wave function $R_{\nu l}(r)$ has the usual analytical form in terms of exponential times Laguerre polynomials. For multi-electron Alkali atoms, $R_{\nu l}(r)$ could for example be numerically found using the Numerov method \cite{book:gallagher}.

 From \eref{re-written_kappa}, using \bref{plane_wave_expansion} and \bref{electronic_wavefunction} we then obtain 
\begin{align} 
\label{kappa_after_plane_wave_expansion}
\kappa^{\nu lm,\nu' l'm'}_{\qv} &=\frac{g_0\sqrt{\rho}}{\sqrt{\cal V}}(\bar{u}_{q}-\bar{v}_{q}) \:\mathcal{N}_{\nu l}\: \mathcal{N}_{\nu' l'} \nonumber \\ 
&\times \int_{0}^{\infty} dr  \:  r^2 R^{*}_{\nu l}(r) R_{\nu' l'}(r) \nonumber \\
& \times \int_{0}^{2\pi} d\Phi \: \int_{0}^{\pi}\:d\Theta \: \text{sin}(\Theta) Y_{lm}^{*}(\Theta,\Phi)Y_{l'm'}(\Theta,\Phi) \nonumber \\
& \times 4 \pi \sum_{l_1,m_1} \:i^{l_1} j_{l_1}(qr) \: Y_{l_1 m_1}(\hat{\mathbf{q}})\: Y_{l_1 m_1}^{*}(\Theta,\Phi).
\end{align} 
The integral over three spherical harmonics gives
\begin{align} 
\label{spherical_harmonic_integral}
&\int_{0}^{2\pi}d\Phi\:\int_{0}^{\pi}d\Theta \:\text{sin}(\Theta)\:Y_{l m}^{*}\:(\Theta,\Phi)\:Y_{l' m'}(\Theta,\Phi)Y_{l_1 m_1}^*(\Theta,\Phi) \nonumber \\
& = (-1)^m(-1)^{m_1}\sqrt{\frac{(2l+1)(2l'+1)(2l_1+1)}{4\pi}}\nonumber \\
&\times\quad
\begin{pmatrix} 
l & l' & l_1 \\
0 & 0 & 0 
\end{pmatrix}
\begin{pmatrix} 
l & l' & l_1 \\
-m & m' & -m_1 
\end{pmatrix},
\quad
\end{align} 
where the last two terms in brackets are Wigner 3-j symbols. We can thus re-write Eq.(\ref{kappa_after_plane_wave_expansion}) as,
\begin{align} 
\label{kappa_after_plane_wave_expansion_1}
\kappa^{\nu lm,\nu' l'm'}_{\qv} &=\frac{g_0\sqrt{\rho}}{\sqrt{\cal V}}(\bar{u}_{q}-\bar{v}_{q}) \mathcal{N}_{\nu l} \mathcal{N}_{\nu' l'} \nonumber \\ 
&\times \int_{0}^{\infty} \text{dr} \:\text{r}^2 R^{*}_{\nu,l}(r) R_{\nu',l'}(r) 4 \pi \sum_{l_1} j_{l_1}(qr) \nonumber \\
& \times \sum_{l_1,m_1}(i)^{l_1}\Bigg[ \sqrt{\frac{(2l+1)(2l'+1)(2l_1+1)}{4\pi}}\: \nonumber \\  
&\times \begin{pmatrix} 
l & l' & l_1 \\
0 & 0 & 0 
\end{pmatrix}
\begin{pmatrix} 
l & l' & l_1 \\
-m & m' & -m_1 
\end{pmatrix} \Bigg]Y_{l_1 m_1}(\hat{\text{q}}),
\end{align}
which will be useful for a general choice of states. Among the experimentally most accessible choices, we now further evaluate \bref{kappa_after_plane_wave_expansion_1} for $s$ ($l=0$, $m=0$) and $p$ ($l=1$, $m=0$) states, within the same 
principal quantum number manifold $\nu=\nu'$.

\subsection{Coupling constant for the Rydberg s-state}
\label{app_coupling_ss_ss} 

For $\alpha=(\nu 00)$ and $\beta=(\nu 00)$ we shall use the shorthand $\kappa^{\nu 00,\nu00}_{\qv}=\kappa^{(ss)}_{\qv}$ for coefficients in \eref{kappa_after_plane_wave_expansion}.
Inserting this choice of quantum numbers into \bref{kappa_after_plane_wave_expansion_1} gives 
\begin{align} 
\label{kappa_ss}
\kappa^{(s s)}_{\qv} =&\frac{g_0\sqrt{\rho}}{\sqrt{\cal V}}(\bar{u}_{q}-\bar{v}_{q}) |\:\Bigg[\Bigg(\frac{2}{\nu a_0}\Bigg)^3 \frac{(\nu-1)!}{2\nu(\nu!)}\Bigg]\nonumber \\
&\times \int_{0}^{\infty} \text{dr} \:\text{r}^2 |R_{\nu 0}(r)|^2 4\pi \sum_{l_1m_1} j_{l_1}(q r)  (i)^{l_1}\nonumber \\
&\times \Bigg[ \sqrt{\frac{(2l_1+1)}{4\pi}}\: 
\begin{pmatrix} 
0 & 0 & l_1 \\
0 & 0 & 0 
\end{pmatrix}
\begin{pmatrix} 
0 & 0 & l_1 \\
0 & 0 & -m_1 
\end{pmatrix}
 \Bigg]Y_{l_1 m_1}(\hat{\mathbf{q}}).
\end{align}
The orthogonality properties encoded in the Wigner-3j symbol now leave only the $l_1=0$, $m_1=0$ term of the double sum, hence
\begin{align} 
\label{kappa_ss_3}
\kappa^{(s s)}_{\qv}=&\kappa^{*(s s)}_{\qv} =\frac{g_0\sqrt{\rho}}{\sqrt{\cal V}}(\bar{u}_{q}-\bar{v}_{q})\Bigg[\Bigg(\frac{2}{\nu a_0}\Bigg)^3 \frac{(\nu-1)!}{2\nu(\nu!)}\Bigg] \CR
&\times  \int_{0}^{\infty} \text{dr} \:\text{r}^2 |R_{\nu 0}(\text{r})|^2j_0(qr),
\end{align}
where we already inserted $Y_{00}(\hat{\mathbf{q}})=1/\sqrt{4\pi}$, and noted that $\kappa^{(s s)}_{\qv}$ is manifestly real. 
The final evaluation of the radial matrix element is deferred to \aref{appendix_radial_ME}.

\subsection{Coupling constant for the Rydberg p-state}
\label{app_coupling_pp_pp} 

Similarly the starting point for $\kappa_{\qv}^{(pp)}$ will be
\begin{align} 
\label{kappa_pp}
\kappa^{(p p)}_{\qv} =&\frac{g_0\sqrt{\rho}}{\sqrt{\cal V}}(\bar{u}_{q}-\bar{v}_{q}) |\:\Bigg[\Bigg(\frac{2}{\nu a_0}\Bigg)^3 \frac{(\nu-2)!}{2\nu[(\nu+1)!]}\Bigg] \nonumber \\
&\times \int_{0}^{\infty} \text{dr} \:\text{r}^2 |R_{\nu 1}(r)|^2 4\pi  \sum_{l_1,m_1} j_{l_1}(qr)(i)^{l_1} \nonumber \\
&\times \Bigg[ \sqrt{\frac{9\:(2l_1+1)}{4\pi}}\:
\begin{pmatrix} 
1 & 1 & l_1 \\
0 & 0 & 0 
\end{pmatrix}
\begin{pmatrix} 
1 & 1 & l_1 \\
0 & 0 & -m_1 
\end{pmatrix}
 \Bigg]Y_{l_1 m_1}(\hat{\text{q}}).
\end{align}
As in case of $\kappa^{(s s)}_{\qv}$ the selection rules in the Wigner symbols help us to restrict the summation for $\kappa^{(pp)}_{\qv}$ in \eref{kappa_pp} to obtain
\begin{align} 
\label{kappa_pp_3}
 \kappa^{(p p)}_{\qv} = &\kappa^{*(p p)}_{\qv}=\frac{g_0\sqrt{\rho}}{\sqrt{\cal V}}(u_{\textbf{q}}-v_{\textbf{q}}) |\:\Bigg[\Bigg(\frac{2}{\nu a_0}\Bigg)^3 \frac{(\nu-2)!}{2\nu[(\nu+1)!]}\Bigg] \nonumber \\
&\times 3 \sqrt{4\pi} \Bigg[ \frac{1}{3}Y_{00}(\hat{\text{q}})\:\int_{0}^{\infty} \text{dr} \:r^2 |R_{\nu 1}(r)|^2j_0(qr) \: \nonumber \\
&- \frac{2\sqrt{5}}{15}Y_{20}(\hat{\text{q}})\:\int_{0}^{\infty} \text{dr} \:r^2 |R_{\nu 1}(r)|^2j_2(qr) \Bigg],
\end{align}
which is again manifestly real.

\subsection{Coupling constant for sp}
\label{app_coupling_sp_sp} 

Finally we follow the same procedure for the coupling constant $\kappa^{(s p)}_{\qv}$ and find
\begin{align} 
\label{kappa_sp_2}
&\kappa^{(s p)}_{\qv} =-\kappa^{*(s p)}_{\qv} =i\:\frac{g_0\sqrt{\rho}}{\sqrt{\cal V}}(\bar{u}_{\text{q}}-\bar{v}_{\text{q}}) \:\:\sqrt{\Bigg(\frac{2}{\nu a_0}\Bigg)^3 \frac{(\nu-1)!}{2\nu(\nu!)}}\:\nonumber \\
& \times \sqrt{\Bigg(\frac{2}{\nu a}\Bigg)^3 \frac{(\nu-2)!}{2\nu[(\nu+1)!]}} \int_{0}^{\infty} \text{dr} \:r^2 R^{*}_{0}(r)R_{1}(\text{r})  \: j_1(qr) Y_{10}(\hat{\text{q}}).
\end{align}
In contrast to the expressions in the two subsections before, this coupling is fully imaginary as indicated.

\subsection{Evaluation of radial matrix elements}
\label{appendix_radial_ME} 
To reach explicit forms for \bref{kappa_ss_3}, \bref{kappa_pp_3} and \bref{kappa_sp_2}, we have to evaluate the remaining radial matrix elements.
\subsubsection{Involving Rydberg $s$-states}
For $\kappa_{\qv}^{(ss)}$, this is possible analytically. From \eref{re-written_kappa} we can write,
\begin{align} 
\label{kappa_ss}
\kappa^{(ss)}_{\qv} &=\frac{g_0\sqrt{\rho}}{\sqrt{\cal V}}(\bar{u}_{q}-\bar{v}_{q}) \int d^3\xv  \:\:  |\psi^{(s)}(\xv)|^2e^{i\qv\cdot\xv}
\end{align} 
When we expand the integration using 3D spherical coordinates defined earlier. The radial part involves an integration over exponential functions times Laguerre polynomials, which has an explicit solution \cite{gradshteyn2014table}. The final coupling constant then takes the form
\begin{align} 
\label{kappa_q_final}
\kappa^{(ss)}_{q} &=\frac{g_0\sqrt{\rho}}{\sqrt{\cal V}}(\bar{u}_{q}-\bar{v}_{q}) \times \frac{i\: \mathcal{N}_{\nu 0}^2}{2q}\times\frac{\Gamma(2\nu) (iq)^{2(\nu-1)}}{(\nu-1)!^2} \nonumber  \\ 
&\times\Bigg[\frac{1}{\Big[\Big(\frac{2}{\nu a_0}\Big)-iq\Big]^{2\nu}}-\frac{1}{\Big[\Big(\frac{2}{\nu a_0}\Big)+iq\Big]^{2\nu}}\Bigg] \nonumber  \\ 
& \times {_2}F_1\Bigg[1-\nu,1-\nu;1-2\nu,1+\left(\frac{2}{\nu a_0 q}\right)^2\Bigg],
\end{align}
where ${_2}F_1$ denotes the Gauss Hypergeometric function and $\Gamma$ the Gamma function.

\subsubsection{Involving Rydberg $p$-states}
Although analytic expressions exist also for radial integrations involving p-states in \bref{kappa_pp_3} and \bref{kappa_sp_2} \cite{shao2013imaging}, these are quite involved and we hence opted to perform those integrations numerically, which will also allow the future incorporation of numerically determined wavefunctions for Alkali atoms. Since we later have to evaluate further integrations over the coupling constant as a function of momenta $\qvec$, it is beneficial to fit the results of the radial numerical integration with a simple functional form, which we describe now. While the coefficient $\kappa^{(ss)}_{q}$ does have an analytical expression in \bref{kappa_q_final}, we found it convenient to treat all coupling on the same footing and also proceed with $\kappa^{(ss)}_{q}$ using the fitting procedure.

\begin{figure}[htb!]
\includegraphics[width=1.0\columnwidth]{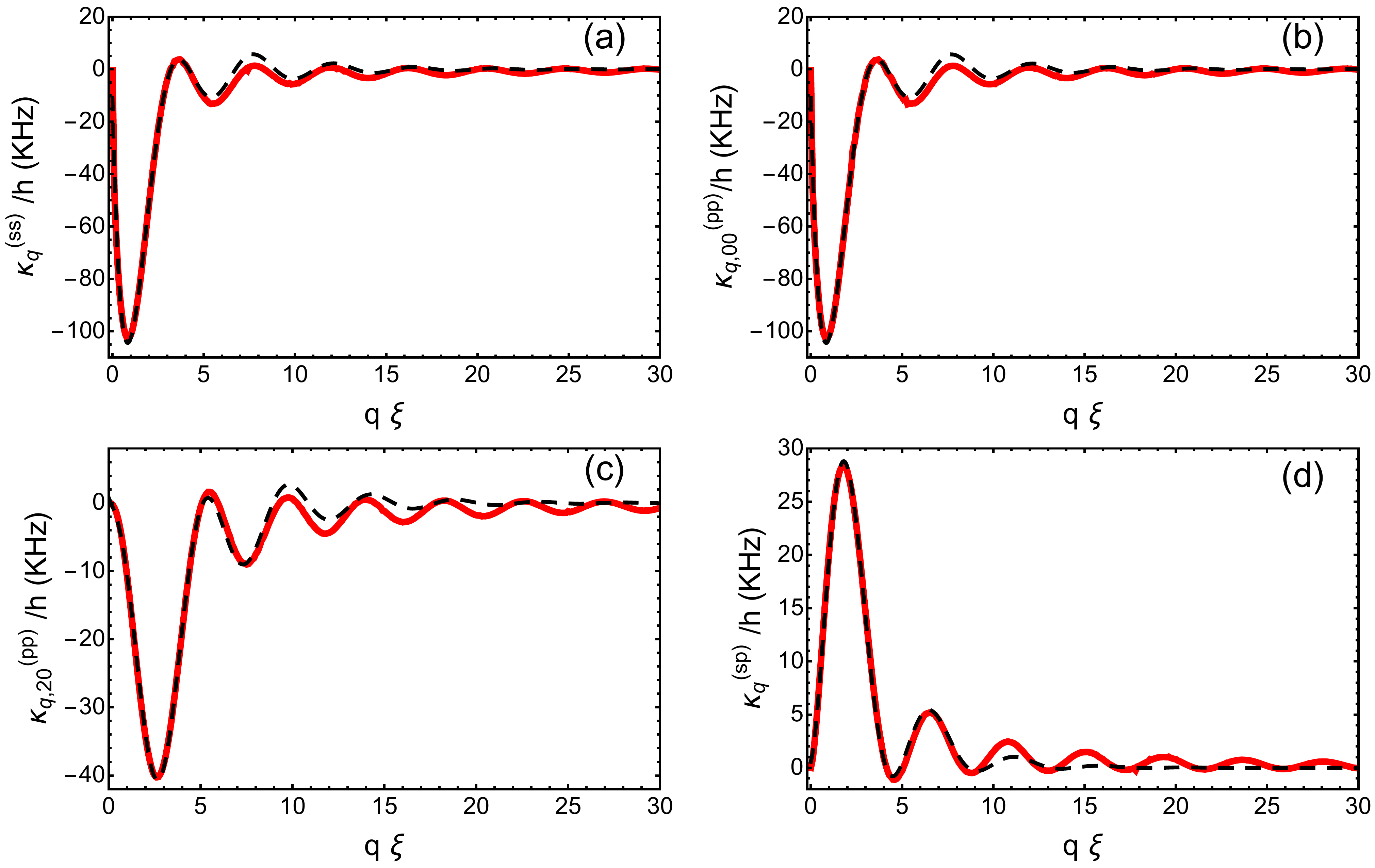}
\caption{\label{fig_fit_function_kappa_pp_sp}  (a) Coupling constant $\kappa^{(ss)}_q$ (red solid) of state $\ket{s}$ to a phonon with wave-number $q$ for principal quantum number $\nu=40$ as in \fref{kappas} and the fit (black-dashed) $g_0f^{(ss)}/\sqrt{\cal{V}}$ from \ref{fit_function_kappa_ss} with the parameters in table \ref{table:fitvalues}.
(b) Isotropic part of the coupling constant $\kappa^{(pp)}_{q,00}$ of state $\ket{p}$ with its the fit $g_0f^{(pp)}_1/\sqrt{\cal{V}}$ and (c) the anisotropic part $\kappa^{(pp)}_{q,20}$ with the fit $g_0f^{(pp)}_2/\sqrt{\cal{V}}$ from \ref{fit_function_kappa_pp}. (d) The transition coupling $\kappa^{(sp)}_q$ between $\ket{s}$ and $\ket{p}$ to a phonon with wave-number $q$ and its fit $g_0f^{(sp)}/\sqrt{\cal{V}}$ from \ref{fit_function_kappa_sp}.} 
\end{figure}

To this end we define a function template 
\begin{align}
\label{fit_function_kappa_template}
T(q)&=A\:\text{sin}(\alpha_1q+\beta_1)e^{-\gamma_1q}+B\:\text{cos}(\alpha_2q+\beta_2)e^{-\gamma_2q},
\end{align}
and then express
\begin{align}
\label{fit_function_kappa_ss}
\sqrt{\cal V}\kappa^{(ss)}_{\qv}/g_0 &= f^{(ss)}(q),\\
\label{fit_function_kappa_sp}
\sqrt{\cal V}\kappa^{(sp)}_{\qv}/g_0 &= f^{(sp)}(q)\cos(\theta),\\
\label{fit_function_kappa_pp}
\sqrt{\cal V}\kappa^{(pp)}_{\qv}/g_0 &= f_1^{(pp)}(q)-f_2^{(pp)}(q)[3\text{cos}^2(\theta)-1],
\end{align}
where each of the functions $f^{(ss)}(q)$, $f^{(sp)}(q)$, $f_1^{(pp)}(q)$, $f_2^{(pp)}(q)$ has the form of the template $T(q)$, with different coefficients $A$, $B$, $\alpha_1$, $\beta_1$, $\gamma_1$, $\alpha_2$, $\beta_2$, $\gamma_2$ as listed in \tref{table:fitvalues}. We have excluded $\sqrt{\cal V}$ from the fit since the quantisation volume must cancel in the calculation of physical quantities later, and $g_0$ to facilitate the conversion of the results to other atomic species.
\begin{table}[htb]
\begin{center}
\begin{tabular}{ |c|c|c|c|c|}
\hline
function & A $\Big[\frac{\mbox{s}}{\mbox{kg m}^{7/2}}\Big]$  & $\alpha_1$ [${\mu}$m] & $\beta_1$ & $\gamma_1$ [${\mu}$m] \\
\hline \hline 
$ f^{(ss)}$ & 2.28$\times10^{45}$ & 0.002 & 0.002 & 0.13  \\ 
 $f_1^{(pp)}$ & 2.19$\times10^{45}$ & 0.002 & 0.002 & 0.135  \\  
 $f_2^{(pp)}$ & 0.54$\times10^{45}$ & -0.002 & -3.15 & 0.07   \\
 $f^{(sp)}$ & 6.94$\times10^{42}$ & 0 & -0.98 & 0.04  \\
 \hline  \hline 
   & B  $\Big[\frac{\mbox{s}}{\mbox{kg m}^{7/2}}\Big]$ & $\alpha_2$ [${\mu}$m] & $\beta_2$ & $\gamma_2$ [${\mu}$m]\\
  \hline  \hline 
  $ f^{(ss)}$ &  -6.82$\times10^{42}$ & 0.16 & 1.48 & 0.025 \\ 
 $f_1^{(pp)}$ &  -6.83$\times10^{42}$ & 0.16 & 1.48 & 0.026 \\  
 $f_2^{(pp)}$ &  -5.84$\times10^{42}$ & 0.15 & -1.02 & 0.024  \\
 $f^{(sp)}$ &  -5.75$\times10^{42}$ & 0.15 & -2.95 & 0.037 \\
  \hline 
\end{tabular}
\end{center}
\label{table:fitvalues}
\caption{Parameters in fit functions \bref{fit_function_kappa_ss}-\bref{fit_function_kappa_pp} for Rydberg-phonon coupling coefficients $\kappa$.
Physical parameters are as for \fref{kappas}. 
}
\end{table}
The quality of these fits is shown in \fref{fig_fit_function_kappa_pp_sp}.
%

\subsection{Formulation of the final Hamiltonian}
\label{final_interaction_Hamiltonian} 

Now that all coupling constants that enter the interaction Hamiltonian \eref{Hamiltonian_matrix_final} in the Rydberg manifolds $\ket{\nu,l=0}$ and $\ket{\nu,l=1}$ are known, via \bref{kappa_ss_3}, \bref{kappa_pp_3} and \bref{kappa_sp_2}, we proceed to re-group that Hamiltonian. Note, that all terms $E^{(\alpha\beta)}$ with $\alpha,\beta\in\{s,p\}$
do not contain BdG mode operators, and hence will be re-allocated to the system Hamiltonian. 
Using the identification $\ket{\uparrow}=\ket{p}$ and $\ket{\downarrow}=\ket{s}$ discussed in \sref{OQS} and Pauli spin matrices, we rewrite the remaining terms as
\begin{align}
\hat{H}_{\text{int}}&=\sum_{\textbf{q}}\frac{\Big(\kappa_{\textbf{q}}^{(pp)}-\kappa_{\textbf{q}}^{(ss)}\Big)}{2}\Big(\hat{b}_{\textbf{q}}+\hat{b}_{\textbf{q}}^{\dagger}\Big) \hat{\sigma}_z \nonumber \\
&+ \sum_{\textbf{q}}\frac{\Big(\kappa_{\textbf{q}}^{(pp)}+\kappa_{\textbf{q}}^{(ss)}\Big)}{2}\Big(\hat{b}_{\textbf{q}}+\hat{b}_{\textbf{q}}^{\dagger}\Big) \mathbbm{1} \nonumber \\
&+i\sum_{\textbf{q}}\kappa_{\textbf{q}}^{(sp)}\Big(\hat{b}_{\textbf{q}}-\hat{b}_{\textbf{q}}^{\dagger}\Big)\hat{\sigma}_y,
\end{align}
where $\mathbbm{1}$ is the unit operator in the Rydberg electronic state space.
Let us define $\bar{\kappa}_{\textbf{q}}=\kappa_{\textbf{q}}^{(pp)}+\kappa_{\textbf{q}}^{(ss)}$ and $\Delta\kappa_{\textbf{q}}=\kappa_{\textbf{q}}^{(pp)}-\kappa_{\textbf{q}}^{(ss)}$ and then consider jointly the present interaction Hamiltonian and the environmental Hamiltonian from \bref{Hamil_g}:
\begin{align}
\label{total_Hamiltonian}
&\sub{\hat{H}}{env}+\sub{\hat{H}}{int}=E_{GP}+\sum_{\qv} \hbar \omega_\qvec \hat{b}^{\dagger}_{\textbf{q}}\hat{b}_{\textbf{q}} +\sum_{\textbf{q}}\frac{\Delta\kappa_{\textbf{q}}}{2}\Big(\hat{b}_{\textbf{q}}+\hat{b}_{\textbf{q}}^{\dagger}\Big) \hat{\sigma}_z \nonumber \\
&+ \sum_{\textbf{q}}\frac{\bar{\kappa}_{\textbf{q}}}{2}\Big(\hat{b}_{\textbf{q}}+\hat{b}_{\textbf{q}}^{\dagger}\Big) \mathbbm{1} +i\sum_{\textbf{q}}\kappa_{\textbf{q}}^{(sp)}\Big(\hat{b}_{\textbf{q}}-\hat{b}_{\textbf{q}}^{\dagger}\Big)\hat{\sigma}_y.
\end{align}
We can then absorb the term $\sim\mathbbm{1}$ by using shifted environmental mode operators
\begin{align}
\label{redefine_Bogoliubov_operators}
\tilde{b}_{\textbf{q}}=\hat{b}_{\textbf{q}}+\frac{\bar{\kappa}_{\textbf{q}}}{2\hbar\omega_q}, \\
\tilde{b}_{\textbf{q}}^{\dagger}=\hat{b}_{\textbf{q}}^{\dagger}+\frac{\bar{\kappa}_{\textbf{q}}^*}{2\hbar\omega_q}.
\end{align} 
We use these in the Hamiltonian \bref{total_Hamiltonian} and then allocate all terms that do not contain environmental operators $\tilde{b}_{\textbf{q}}$ or $\tilde{b}^{\dagger}_{\textbf{q}}$  to a shifted system Hamiltonian,  so that our final result for the complete Hamiltonian $\hat{H}=\sub{\hat{H}}{syst}' + \sub{\hat{H}}{coup} + \sub{\hat{H}}{env}' +const$ becomes
\begin{subequations}
\label{final_hamil_rearranged}
\begin{align}
\label{final_Hsyst}
\sub{\hat{H}}{syst}' &=\sub{\hat{H}}{syst}  -\sum_{\qv}\frac{\Delta\kappa_{\textbf{q}}\bar{\kappa}_{\textbf{q}}}{2\hbar\omega_q} \hat{\sigma}_z\\ 
\label{final_Henv}
\sub{\hat{H}}{env}' &=\sum_{\qv}\hbar\omega_q\:\tilde{b}^{\dagger}_{\qv}\tilde{b}_{\textbf{q}} ,\\
\label{final_Hcoup}
 \sub{\hat{H}}{coup}&=\sum_{\qv} \frac{\Delta\kappa_{\textbf{q}}}{2}\Big(\tilde{b}_{\textbf{q}}+\tilde{b}_{\textbf{q}}^{\dagger}\Big)\hat{\sigma}_z \CR
 &+i\sum_{\textbf{q}}\kappa_{\textbf{q}}^{(sp)}\Big(\tilde{b}_{\textbf{q}}-\tilde{b}_{\textbf{q}}^{\dagger}\Big)\hat{\sigma}_y,
\end{align}
\end{subequations}
where the constant energy offset in $\hat{H}$ has absorbed some contributions from \bref{total_Hamiltonian}. 
 
We can evaluate the environment induced energy shift in $H'_{syst}$ after converting the discrete summation over modes to a continuous integral, $\sum_{\qv}\longrightarrow\int d^3{\qv}D$, with density of states $D={\cal V}/(2\pi)^3$ and find
\begin{align}
\label{H_syst_integral}
 &\bar{E}\hat{\sigma}_z\equiv -\sum_{\qv}\frac{\Delta\kappa_{\textbf{q}}\bar{\kappa}_{\textbf{q}}}{2\hbar\omega_q} \hat{\sigma}_z=
D\int d^3{\qv} \frac{\Delta\kappa_{\textbf{q}}\bar{\kappa}_{\textbf{q}}}{2\hbar\omega_q{\cal V}}\hat{\sigma}_z=\CR
&2\pi D \int_0^{\infty}dq\:q^2\Big[\frac{2 f_1^{(pp)}(q)^2+\frac{8}{5}f_2^{(pp)}(q)^2-2 f^{(ss)}(q)^2}{2\hbar\omega_q}\Big]\hat{\sigma}_z,
\end{align}
where we already integrated over $\theta$ and $\varphi$. Evaluating the final integral we reach for example $\bar{E}=19$ GHz using the parameters of \fref{kappas}.
For later use, we finally split $ \sub{\hat{H}}{coup}=\hat{\sigma}_z \otimes \hat{E}^{(z)}+\hat{\sigma}_y \otimes \hat{E}^{(yy)}$, with 
\begin{align}
\label{z_environmental_operator}
\hat{E}^{(z)}&=\sum_{\mathbf{q}}\frac{\kappa^{(pp)}_{\textbf{q}}-\kappa^{(ss)}_{\textbf{q}}}{2}\Big[\tilde{b}_{\textbf{q}}+\tilde{b}_{\textbf{q}}^{\dagger}\Big],\\
\label{y_environmental_operator}
\hat{E}^{(yy)}&=i\sum_{\mathbf{q}}\kappa^{(sp)}_{\textbf{q}}\Big[\tilde{b}_{\textbf{q}}-\tilde{b}_{\textbf{q}}^{\dagger}\Big].
\end{align}

A very important final point, is that after re-defining the BdG operators as in \bref{redefine_Bogoliubov_operators}, they fulfill
\begin{align}
\label{New_Bogiluibov_operator_on_vacuum_state_1}
\tilde{b}_{\textbf{q}}\ket{0}=d_{\textbf{q}}\ket{0},
\\
\label{New_Bogiluibov_operator_on_vacuum_state_2}
\bra{0}\tilde{b}^{\dagger}_{\textbf{q}}=\bra{0}d^*_{\textbf{q}},
\end{align}
for $d_{\textbf{q}}=\frac{\bar{\kappa}_{\qvec}}{2\hbar\omega_q}$, where $\ket{0}$ is the BdG vacuum for the original unshifted operators $\hat{b}$. These equations make clear, that in terms of the new operators, the original BdG vacuum is a many-mode coherent state or \emph{displaced} vacuum. However, we will show in the next section, that open quantum system dynamics with an environment initialised in a coherent state is equivalent to one with an environment in a vacuum state  and a slight shift in the Hamiltonian. Hence we subsequently consider also the environment state $\hat{\rho}_{\cal E}$ for the newly defined operators $\tilde{b}_{\textbf{q}}$ to be the vacuum state.

\section{Transformation of environmental state}
\label{app_coherentstate_treatment} 
%
We had seen in the preceding appendix, that in order to reach a standard form of the Spin-Boson model in \ref{final_hamil_rearranged}, it has to be formulated in terms of Bogoliubov operators shifted as in \eref{redefine_Bogoliubov_operators}. Thus, the initial environment vacuum state $\hat{\rho}_{\cal E}=\ket{0}\bra{0}$ for operators $\hat{b}_\mathbf{q}$ becomes  $\hat{\rho}_{\cal E}=\ket{\dv}\bra{\dv}$ for operators $\tilde{b}_{\textbf{q}}$.

Let us insert this state into \bref{formal_reddm_evolution}, and then rewrite the many mode coherent state using the standard displacement operator $\ket{\dv}=\hat{D}(\dv)\ket{0}$. We find
\begin{align}
\label{formal_reddm_evolution_step2}
&\hat{\rho}_{\cal S}(t) =\mbox{Tr}_{\cal E} \left(\hat{U}(t)   \left[\hat{\rho}_{\cal S}(0)\otimes \ket{\dv}\bra{\dv}\right] \hat{U}^\dagger(t)\right)\\
&=\mbox{Tr}_{\cal E} \left(\hat{U}(t)   \left[\hat{\rho}_{\cal S}(0)\otimes\hat{D}(\dv) \ket{0}\bra{0}\hat{D}^\dagger(\dv)\right] \hat{U}^\dagger(t) \hat{D}(\dv) \hat{D}^\dagger(\dv)\right), \nonumber
\end{align}
where we have also inserted $\mathbbm{1} =  \hat{D}(\dv) \hat{D}^\dagger(\dv)$ into the trace.

This can be re-arranged into
\begin{align}
\label{formal_reddm_evolution_step3}
&\hat{\rho}_{\cal S}(t) =\mbox{Tr}_{\cal E} \left(\tilde{U}(t)   \left[\hat{\rho}_{\cal S}(0)\otimes \ket{0}\bra{0}\right] \tilde{U}^\dagger(t)\right),
\end{align}
where the time-evolution is now governed by the shifted time evolution operator $\tilde{U}(t) =  \hat{D}(\dv) \hat{U}(t) \hat{D}^\dagger(\dv)$, but starts from a vacuum environment initial state. That time evolution operator arises in turn from a shifted Hamiltonian $\tilde{H}= \hat{D}(\dv) \hat{H}(t) \hat{D}^\dagger(\dv)$, where $\hat{H}(t)$ is the interaction picture Hamiltonian following from \bref{final_hamil_rearranged}.

This shifted Hamiltonian finally takes the form 
\begin{align}
\label{shift_piece_hamil}
\sub{\hat{H}}{syst}''&=\sub{\hat{H}}{syst}'+\hat{\sigma}_z \sum_{\qv}\frac{\Delta\kappa_{\qv}\:\bar{\kappa}_{\qv}}{2\hbar\omega_q}\:\text{cos}(\omega_qt),
\end{align}
where $\sub{\hat{H}}{syst}'$ was given in \eref{final_hamil_rearranged}. We shall explore the ramifications of this term for our scenario in \cite{rammohan:superpos}.
A similar procedure was used in \cite{hartmann_hops_JCTC} to handle finite temperature environments.

\section{Calculation of BEC environment correlation functions}
\label{app_correlfct} 

As discussed in \sref{correlfct}, the effect of the BEC environment on the Rydberg impurity is fully encapsulated in the environmental correlation functions defined in  
\eref{C_of_tau_vac}. These equations define three different correlation functions, owing to the two non-trivial parts of the system-environment coupling Hamiltonian \bref{final_Hcoup}. Correlations depend on the assumed state of the environment, for which we can take the vacuum state as shown in \sref{app_coherentstate_treatment}.

\subsection{zz Correlations}
\label{app_zz_correlfct} 
%
Inserting \eref{z_environmental_operator} into $C^{(zz)}(\tau)$ of \eref{C_of_tau_vac} we can write
\begin{align}
\label{Correlation_our_system_zz}
&C^{(zz)}(\tau)=\sum_{\textbf{q},\mathbf{q'}}\frac{\Delta\kappa_{\mathbf{q}}}{2}\frac{\Delta\kappa_{\mathbf{q'}}}{2}\nonumber \\
&\times \bra{0}\Big[\tilde{b}_{\textbf{q}}(\tau)+\tilde{b}_{\textbf{q}}^{\dagger}(\tau)\Big]\Big[\tilde{b}_{\mathbf{q'}}(0)+\tilde{b}_{\mathbf{q'}}^{\dagger}(0)\Big]\ket{0}.
\end{align}
In the interaction picture we have 
\begin{align}
\label{Interaction_b_bdagger}
&\tilde{b}_{\textbf{q}}(\tau)=\tilde{b}_{\textbf{q}}(0)e^{-i\omega_q\tau}, \\
&\tilde{b}^{\dagger}_{\textbf{q}}(\tau)=\tilde{b}^{\dagger}_{\textbf{q}}(0)e^{i\omega_q\tau}.
\end{align}
Hence \eref{Correlation_our_system_zz} becomes,
\begin{align}
\label{Correlation_zz}
C^{(zz)}(\tau)=\sum_{\mathbf{q}}\frac{\Delta\kappa_{\mathbf{q}}^2}{4}e^{-i\omega_q\tau}.
\end{align}
To evaluate \bref{Correlation_zz}, we again convert from the discrete to a continuous notation, according to $\sum_{\qv}\longrightarrow\int d^3 \qv\: D$. As expected, we see that the quantisation volume ${\cal V}$ from the $\sum_{\qv}$ cancels, those from $\Delta\kappa_\qvec$ and $d_\qvec$, see e.g.~\eref{kappa_ss_3}. Let us denote the 3D spherical coordinates of the wave-vector $\qvec$ with $q=|\qvec|$, $\theta$ and $\varphi$. We now insert the fitted coupling constants obtained in \aref{appendix_radial_ME} to obtain
\begin{align}
\label{Alphakappa_equal_fit_function}
d_{\qv}&=\frac{g_0}{\sqrt{\cal V}}\frac{ f_1^{(pp)}(q) -f_2^{(pp)}(q) (3\text{cos}^2(\theta)-1)+f^{(ss)}(q)}{2\hbar\omega_q},\\
\label{Deltakappa_equal_fit_function}
\Delta\kappa_{\qv}&=\frac{g_0}{\sqrt{\cal V}}\left[  f_1^{(pp)}(q) -f_2^{(pp)}(q) (3\text{cos}^2(\theta)-1)-f^{(ss)}(q) \right].
\end{align}
 Evaluating angular integrals we find
\begin{align}
\label{correlation_after_putting_fit_function_zz_2}
&C^{(zz)}(\tau)=\pi g_0^2D \int_{0}^{\infty}dq\:q^2e^{-i\omega_{\mathbf{q}}\tau} \times \bigg[ f_1^{(pp)}(q)^2+\frac{4}{5}f_2^{(pp)}(q)^2 \CR
&+f^{(ss)}(q)^2-2f_1^{(pp)}(q)f^{(ss)}(q) \bigg],
\end{align}
%
after also integrating over the azimuthal angle $\varphi$.
The integrations over $q$ are finally performed numerically, with results shown in \fref{fig_correlation_functions}.

\subsection{yy Correlations}
\label{app_correlation_yy} 
%
The calculation of environmental correlation functions involving the operator $\hat{E}^{(yy)}$ proceeds similarly.
After insertion of interaction picture bath operators, we now have 
\begin{align}
\label{Correlation_our_system_sp_2}
C^{(yy)}(\tau)=&\sum_{\qvec}\left[\kappa_\qvec^{(sp)}\right]^2 \bra{0} \tilde{b}_\qvec(0)e^{-i\omega_\qvec \tau}\tilde{b}_\qvec^{\dagger}(0)\ket{0}.
\end{align}
After the same conversion from discrete to continuous bath modes as in the previous section, we reach
\begin{align}
\label{Correlation_yy}
 C^{(yy)}(\tau)&=-\frac{4\pi}{3} g_0^2 D \int_{0}^{\infty} dq\:q^2 f^{(sp)}(q)^2e^{-i\omega_q\tau}.
\end{align}
%
\subsection{yz correlations}
\label{app_correlation_zzyy} 
%
For a system environment Hamiltonian containing two coupling terms such as \bref{final_Hcoup}, in principle also cross correlation functions between environmental operators in those two terms may become relevant. However we show now that 
 \begin{equation} 
C^{(zy)}(\tau)=\bra{0} \hat{E}^{(z)}(\tau) \hat{E}^{(y)}(0) \ket{0},
 \label{Ccross_of_tau_coh}
 \end{equation} 
vanishes in our case. As before we insert \bref{z_environmental_operator} and \bref{y_environmental_operator} into \bref{Ccross_of_tau_coh} to find
\begin{align}
\label{Correlation_applying_coherent_property_zz_yy_3}
C^{(zy)}(\tau)=-i\sum_{\mathbf{q}}\frac{\Delta\kappa_{\mathbf{q}}}{2}\kappa^{(sp)}_{\mathbf{q}}e^{-i\omega_{\mathbf{q}}\tau}.
\end{align} 
The angular structure of the resultant integral is odd and the integral vanishes, so that $C^{(zy)}(\tau)=0$.

\section{Calculation of spectral densities}
\label{app_specdens} 

As discussed in \sref{spec_dens_tuning}, spectral densities contain interesting information on environmental properties.
We can obtain them directly from the definition \bref{specden}
\begin{equation} 
J^{(z)}(\omega) =\sum_\qvec\frac{\Delta\kappa_{\qvec}^2}{4}\delta(\omega -\omega_\qvec)
 \label{specden_app}
\end{equation} 
If we convert the sum to a continuum integral we can write this as
\begin{align}
\label{specden_app_cont}
J^{(z)}(\omega) = \int_0 ^ {\infty} d^3\qv \: \frac{\Delta\kappa_{\qvec}^2}{4}\delta(\omega -\omega_\qvec) D,
\end{align}
 where $\omega_{\qvec}=\sqrt{\frac{\hbar q^2}{2m}\Big(\frac{\hbar^2q^2}{2m}+2U_0\rho\Big)}$. To evaluate the delta-function
 we require
\begin{align}
\frac{d\omega_q}{dq}=\frac{\frac{\hbar^2q^2}{m^2}+\frac{2U_0\rho}{m}}{2\sqrt{\frac{\hbar^2q^2}{4m^2}+\frac{U_0\rho}{m}}}
\end{align}
and using the parametrisation \bref{fit_function_kappa_ss}, \bref{fit_function_kappa_pp} for $\Delta\kappa_{\qvec}$ we finally reach: 
\begin{align}
\label{spec_dens_sum_final_1}
J^{(z)}(\omega)=&\frac{\pi D}{\sqrt{2}}   q_{\omega}^2\Big[f_1^{(pp)}(q_{\omega})^2+\frac{4}{5}f^{(ss)}(q_{\omega})^2\nonumber \\
&+f^{(sp)}(q_{\omega})^2-2f_1^{(pp)}(q_{\omega})f^{(ss)}(q_{\omega})\Big]\frac{dq_\omega}{d\omega_q},
\end{align}
%
with $q_{\omega}=\frac{\sqrt{2m}}{\hbar}\:\sqrt{\sqrt{(\hbar\omega)^2+(U_0\rho)^2}-U_0\rho}$.

Similarly the spectral density for the $\hat{\sigma}_y$ coupling can be written as,
\begin{align}
\label{sigma_y_specdens}
J^{(y)}(\omega) = \int_0 ^ {\infty} d^3\qv \: \kappa_{\qvec}^{(sp)\:2}\delta(\omega -\omega_\qvec) D,
\end{align}
which will have the final form,
\begin{align}
\label{spec_dens_y_sum_final}
J^{(y)}(\omega)=&\frac{\pi  D}{\sqrt{2}}  q_{\omega}^2\Bigg[\frac{1}{3}f^{(sp)}(q_{\omega})^2\Bigg]\frac{dq_\omega}{d\omega_q}.
\end{align}
We explicitly verified that the same spectral densities are obtained via the Fourier transform relation \bref{JfromC}.


%

\end{document}